\documentclass[11pt,review]{elsarticle}
\usepackage{ucs} 
\usepackage[utf8x]{inputenc} 
\usepackage[T1]{fontenc}
\usepackage{lineno,hyperref}
\modulolinenumbers[5]
\journal{arXiv}
\hypersetup{
    colorlinks,
    linkcolor={red!50!black},
    citecolor={blue!50!black},
    urlcolor={blue!80!black}
}

\usepackage[bitstream-charter]{mathdesign}
\DeclareSymbolFont{usualmathcal}{OMS}{cmsy}{m}{n}
\DeclareSymbolFontAlphabet{\mathcal}{usualmathcal}

\usepackage[superscript]{cite}

\usepackage{bm,amsmath,mathrsfs,graphicx,psfrag}
\usepackage{pst-all,epsfig,wrapfig,subfigure,rotating} 
\usepackage[final,inline]{showlabels} 
\usepackage{mdframed,enumitem}
\usepackage{ifpdf}
\usepackage[normalem]{ulem}
\usepackage{xcolor}
\usepackage{sectsty}
\usepackage{tikz}
\usetikzlibrary{calc,arrows,decorations.pathreplacing,decorations.markings}
\sectionfont{\normalsize}

\renewcommand{\strut}{\rule{0pt}{15pt}}
\newcommand{\titlestrut}{\rule{0pt}{25pt}}
\newcommand{\ii}{\text{i}}
\renewcommand{\i}{\text{i}}
\newcommand{\del}{\partial}
\newcommand{\bdel}{\bar\partial}

\newcommand{\half}{\frac{1}{2}}

\newcommand{\bs}[1]{\boldsymbol{#1}}
\renewcommand{\d}{\text{d}}
\newcommand{\x}{\text{x}}
\newcommand{\y}{\text{y}}
\newcommand{\z}{\text{z}}

\newcommand{\comm}[2]{\left[#1,#2\right]}

\newcommand{\anticomm}[2]{\left\{#1,#2\right\}}

\newcommand{\parder}[2]{\frac{\partial #1}{\partial #2}}
\newcommand{\abs}[1]{\left|#1\right|}
\newcommand{\up}{\uparrow}
\newcommand{\dw}{\downarrow}

\newcommand{\ket}[1]{\left|#1\right\rangle}

\newcommand{\vac}{\left|0\right\rangle}

\def\ie{{i.e.},\ }
\def\eg{{e.g.}\ }
\def\etal{{et al.}}

\def\viz{{viz.}\ }
\def\s{\scriptscriptstyle}

\let\OLDcaption\caption
\renewcommand\caption[1]{\OLDcaption{\baselineskip12pt #1}}
\newcommand{\cancel}[1]{\textcolor{red}{#1}}
\renewcommand{\cancel}[1]{}

\RequirePackage[top=20mm,bottom=31mm,left=36mm,right=34mm,head=2mm,includeheadfoot]{geometry}
\usepackage{titlesec}
\titleformat*{\subsection}{\itshape}
\begin{document}

\begin{frontmatter}
\setlength{\baselineskip}{14.8pt}
\title{\titlestrut {\bf Fractional Statistics}\\
}

\author{Martin Greiter} 
\address{Institut für Theoretische Physik und Astrophysik and Würzburg-Dresden Cluster of Excellence ct.qmat, Julius-Maximilians-Universität, 97074 Würzburg, Germany \\[-125pt]
\hfill\fbox{\rm\footnotesize MIT-CTP/5401}\\[110pt]
{\rm greiter@physik.uni-wuerzburg.de}}
\author{Frank Wilczek}
\address{Center for Theoretical Physics, MIT, Cambridge, MA 02139 USA; \\
T.D.~Lee Institute and Wilczek Quantum Center, 
Shanghai Jiao Tong University, Shanghai 200240, China;\\
Arizona State University, Tempe, AZ 85287, USA; \\
Fysikum, Stockholm University, Stockholm, Sweden\\[\smallskipamount] 
{\rm wilczek@mit.edu}\vspace{-30pt}}

\begin{abstract} 
\setlength{\baselineskip}{14pt}
The quantum-mechanical description of assemblies of particles whose motion is confined to two (or one) spatial dimensions offers many possibilities that are distinct from bosons and fermions.  We call such particles anyons.  The simplest anyons are parameterized by an angular phase parameter $\theta$ (i.e., a complex number $e^{i \theta}$ of unit magnitude).  $\theta = 0, \pi$ correspond to bosons and fermions respectively; at intermediate values we say that we have fractional statistics.  In two dimensions, $\theta$ describes the phase acquired by the wave function as two anyons wind around one another counterclockwise.  It generates a shift in the allowed values for the relative angular momentum.  Composites of localized electric charge and magnetic flux associated with an Abelian U(1) gauge group (or a discrete subgroup thereof) realize this behavior.  More complex charge-flux constructions can involve non-Abelian and product groups acting on a spectrum of allowed charges and fluxes, giving rise to non-Abelian and mutual statistics.  Interchanges of non-Abelian anyons implement unitary transformations of the wave function within an emergent space of internal states.  Mutual statistics associates non-trivial transformations with winding of distinguishable particles.  Anyons of all kinds are described by quantum field theories that include Chern--Simons terms.  The crossings of one-dimensional anyons on a ring are uni-directional, such that a fractional phase $\theta$ acquired upon interchange gives rise to fractional shifts in the relative momenta between the anyons.  The quasiparticle excitations of fractional quantum Hall states have long been predicted to include anyons.  Recently the anyon behavior predicted for quasiparticles in the $\nu = 1/3$ fractional quantum Hall state has been observed both in scattering and in interferometric experiments.  The latter demonstrate the defining feature of anyons, \ie phase accumulation by braiding, quite directly.  Anyons are also predicted to occur in spin liquids and other hypothetical states of matter.  Excitations within designed systems, notably including superconducting circuits, can exhibit anyon behavior.  Such systems are being developed for possible use in quantum information processing.  
\end{abstract}

\end{frontmatter}
\setlength{\baselineskip}{15pt}
\subsection*{Keywords}

Anyons, braid group, charge-flux tube composites,  Chern--Simons term, fractional exclusion principle, fractional momentum spacings, fractional relative angular momentum, Haldane--Shastry model, Ising anyons, Laughlin state, Moore--Read state, non-Abelian statistics, Pfaffian state.

\subsection*{Key Points/Objectives}
\begin{itemize}
\item Assemblies of quantum particles whose motion is confined to two spatial dimensions, or to one, can exhibit behavior that differs from fermions and bosons, and can interpolate between them.  Such particles are called anyons, and are said to obey fractional statistics.
\item Fractional statistics is associated with fractional relative angular momentum, and thereby with characteristic centrifugal barriers.
\item In one dimension, crossings of anyons are uni-directional, and the momentum spacings are fractionally quantized. 
\item The simplest anyon behavior is realized by charge-flux tube composites, where the charge and flux are associated with an Abelian gauge group.  
\item More complex anyon theories can feature non-Abelian groups and a spectrum of charges and fluxes.  These realize non-Abelian and mutual statistics.
\item Non-Abelian anyons correspond to higher dimensional representations of the braid group.  Interchanges alter the state vector in an internal Hilbert space supported by the anyons. 
\item Anyons are described by quantum field theories that include a Chern-Simons term.
\item The quasiparticle excitations of fractionally quantized Hall states are predicted to be anyons.
\item Abelian anyon behavior has been observed in fractionally quantized Hall states, through both scattering and interferometric experiments.  The latter exhibit the central phenomenon of phase accumulation by braiding quite directly.
\item Excitations within designed systems, including superconducting circuits, can exhibit anyon behavior.  These realizations are being developed for possible use in quantum information processing.
\item p-wave superfluids, and the Moore--Read state, are predicted to support the simplest non-Abelian anyons, the Ising anyons.  Their behavior can be understood in terms of Majorana fermion modes at the vortex cores.
\end{itemize}


\section{Introduction}
For many decades after the discovery of quantum physics early last century, there was a universally shared belief that identical particles would be either fermions or bosons.  The simplest explanation was that since the many-particle Hamiltonians are by definition symmetric under interchanges of particles, the wave functions have to transform under a representation of the permutation group $S_N$.  Since the wave functions are complex valued scalars, these representations would have to be one dimensional, and the permutation group has only two such representations, the antisymmetric and the symmetric one.  Particles described by wave functions transforming under either representation are called fermions and bosons.

Another angle of approach is through a deep connection between the intrinsic angular momentum or spin of particles, and their mutual quantum statistics in quantum field theory, called the spin-statistics theorem\cite{finkelstein-68jmp1762}.  It states 
that fermions have half-integer spin, while bosons have integer spin.  Since the generators of rotations in three spatial dimensions (3D), the components of angular momentum or spin, obey the su(2) commutation relations 
\begin{align}
 \label{eq:Lcomm}
  \comm{S_i}{S_j}=\ii\hbar\epsilon_{ijk}S_k,
\end{align}
where $\epsilon_{ijk}$ is the totally antisymmetric tensor, and the allowed values $s$ for the quantization of angular momentum or spin according to $\bs{S}^2=\hbar^2 s(s+1)$ are limited to $s=0,\frac{1}{2},1,\frac{3}{2},2,\ldots$.  Half integer values for $s$ correspond to fermions, while integer values correspond to bosons.  Since \eqref{eq:Lcomm} precludes fractional values for $2s$, the only allowed choices for the statistics are once again fermions and bosons.

The latter considerations, however, only apply to 3D.  In two spatial dimensions (2D), there is only one generator of angular momentum (which generates rotations within the plane), and spin does not have to be quantized.  The spin-statistics theorem hence does not exclude the possibility of exotic quantum statistics interpolating between bosons and fermions\cite{Wilczek90}.  At first glance, however, the argument relying on the representations of the permutation group appears to hold in 2D as well.

\section{Path integrals and the braid group}

\begin{figure}[bt] 
  \begin{center}
    \pgfmathsetmacro{\rad}{.5}
    \begin{tikzpicture}[>=latex,scale=.8]
      \begin{scope}[shift={(0,0)}]
      \draw [fill,thick]( 0,0) circle[radius=.04];
      \draw [fill,thick]( 1,0) circle[radius=.04];
      \draw [blue,thick](1,0) to [out=90,in=0] (0,1);
      \draw[blue,thick,->] (0,1) to [out=180,in=90] (-1,-0.06);
      \node at (0.2,-1) {$e^{\ii\theta}$};
      \end{scope}
      \begin{scope}[shift={(4,0)}]
      \draw [fill,thick]( 0,0) circle[radius=.04];
      \draw [fill,thick]( 1,0) circle[radius=.04];
      \draw [blue,thick](1,0) to [out=90,in=0] (-.1,1.1);
      \draw [blue,thick](-.1,1.1) to [out=180,in=90] (-1.2,0);
      \draw [blue,thick](-1.2,0) to [out=270,in=180] (-.2,-1);
      \draw [blue,thick](-.2,-1) to [out=0,in=270] (.8,0);
      \draw [blue,thick](.8,0) to [out=90,in=0] (-.1,.9);
      \draw [blue,thick,->] (-.1,.9) to [out=180,in=90] (-1,-.06);
      \node at (1.2,-1) {$e^{3\ii\theta}$};
      \end{scope}
      \begin{scope}[shift={(7.5,0)}]
      \draw [fill,thick]( 0,0) circle[radius=.04];
      \draw [fill,thick]( 1,0) circle[radius=.04];
      \draw [blue,thick] (1,0) to [out=30,in=270] (1.4,.6);
      \draw [blue,thick] (1.4,.6) to [out=90,in=0] (1,1);
      \draw [blue,thick] (1,1) to [out=180,in=90] (.6,.6);
      \draw [blue,thick,->] (.6,.6) to [out=270,in=135] (.94,.04);
      \node at (1,-1) {$1$};
      \end{scope}
      \begin{scope}[shift={(11.5,0)},yscale=-1]
      \draw [fill,thick]( 0,0) circle[radius=.04];
      \draw [fill,thick]( 1,0) circle[radius=.04];
      \draw [blue,thick](1,0) to [out=90,in=0] (0,1);
      \draw [blue,thick](0,1) to [out=180,in=90] (-1,0);
      \draw [blue,thick](-1,0) to [out=270,in=180] (0,-1);
      \draw [blue,thick,->] (0,-1) to [out=0,in=262] (.99,-.07);
      \end{scope}
      \begin{scope}[shift={(11.5,0)},yscale=1]
      \node at (1.2,-1) {$e^{-2\ii\theta}$};
      \end{scope}
    \end{tikzpicture}
  \end{center}
  \caption{Examples of topologically distinct windings of two particles in 2D with and without interchanges, and the associated phase factors according to \eqref{eq:tau1D}.}
  \label{fig:2D_windings} 
\end{figure}
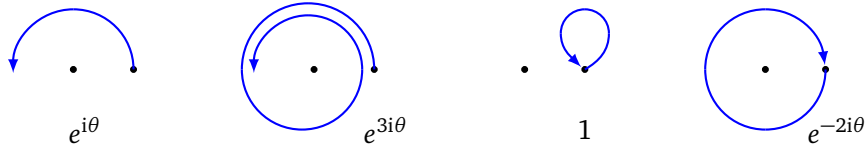

This consensus was challenged by Leinaas and Myrheim\cite{leinaas-77ncb1} in 1977 (see also Goldin \etal\cite{goldin-81jmp1664}).  They pointed out that the fundamental quantity to consider was not the wave function, but the relative amplitudes of paths belonging to topological distinct sectors when particles are interchanged, which are constrained only by unitarity and the composition principle\cite{FeynmanHibbs65}.  In 2D, we can define winding numbers as we interchange particles, and all paths with different windings are topologically distinct (see Figure \ref{fig:2D_windings}).  In 3D, by contrast, the only topological distinct sectors are those where particles are interchanged or not interchanged, as all windings can be untangled.

The relevant group in 2D, therefore, is not permutation group $S_N$, but the braid group $B_N$.  This is the group of all possible ways to interchange $N$ particles through windings, and is generated by the $N-1$ counterclockwise interchanges $T_i$ of particles labeled by consecutive indices $i$ and $i+1$.  The algebra of the group is given by\cite{Kauffman93}
\begin{align}
  \label{eq:pfBraidAlgebra}
  \begin{array}{r@{\ }c@{\ }ll}
    T_iT_j &=& T_jT_i      \quad&\text{for}\ \,|i-j|>1,\\ 
    T_iT_jT_i &=& T_jT_iT_j\quad&\text{for}\ \,|i-j|=1,
  \end{array}
\end{align}
as illustrated in Figure \ref{fig:pfBraidingDef}.  Note that the braid group differs from the permutation group as $T_i^{-1}\ne T_i$. The one dimensional representations of $B_N$ on $\mathbb{R}^2$ are given by 
\begin{align}
  \label{eq:tau1D}
  \tau(T_i)=e^{\ii\theta},
\end{align}
and hence labeled by a continuous U(1) phase parameter $\theta\in\ ]-\pi,\pi]$.

\newcommand{\braiding}
   {\draw[thin]( 0,0) arc[start angle=0,end angle=54,radius=1];
    \draw[thin](-1,0) arc[start angle=180,end angle=120,radius=1];
    \draw[thin]( 0,{sqrt(3)}) arc[start angle=0,end angle=-60,radius=1];
    \draw[thin](-1,{sqrt(3)}) arc[start angle=-180,end angle=-126,radius=1];}
\newcommand{\nodes}
   {\node[below] at(-1,0){\small $\gamma_{i}$};
    \node[below] at (0,0){\small $\gamma_{i+1}$};
    \node[below] at (1,0){\small $\gamma_{i+2}$};}
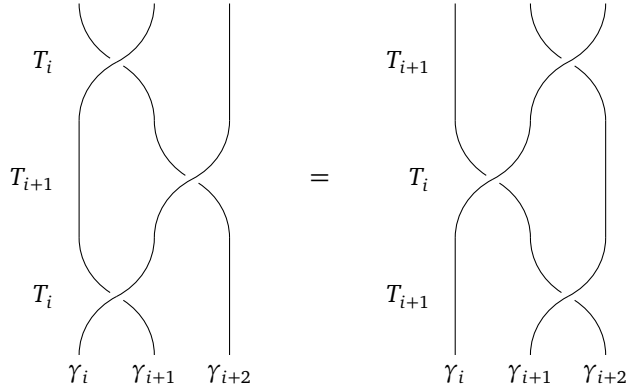
\begin{figure}[bt] 
  \vspace{15pt}
  \begin{center}
  \pgfmathsetmacro{\rad}{.5}
  \begin{tikzpicture}[>=latex,xscale=1,yscale=.9]
  \begin{scope}[shift={(0,0)}]
    \begin{scope}[shift={(0,{2*sqrt(3)})}]
      \braiding
      \draw[thin](1,0)--(1,{sqrt(3)});
      \node[left] at (-1.2,{sqrt(3)/2}){\small $T_{i}$};  
    \end{scope}
    \begin{scope}[shift={(1,{sqrt(3)})}]
      \braiding
      \draw[thin](-2,0)--(-2,{sqrt(3)});
      \node[left] at (-2.2,{sqrt(3)/2}){\small $T_{i+1}$};  
    \end{scope}
    \begin{scope}[shift={(0,0)}]
      \braiding
      \draw[thin](1,0)--(1,{sqrt(3)});
      \node[left] at (-1.2,{sqrt(3)/2}){\small $T_{i}$};  
    \end{scope}
    \nodes
  \end{scope}
  \node at (2.2,{3*sqrt(3)/2}){\small $=$};  
  \begin{scope}[shift={(5,0)}]
    \begin{scope}[shift={(1,{2*sqrt(3)})}]
      \braiding
      \draw[thin](-2,0)--(-2,{sqrt(3)});
      \node[left] at (-2.2,{sqrt(3)/2}){\small $T_{i+1}$};  
    \end{scope}
    \begin{scope}[shift={(0,{sqrt(3)})}]
      \braiding
      \draw[thin](1,0)--(1,{sqrt(3)});
      \node[left] at (-1.2,{sqrt(3)/2}){\small $T_{i}$};  
    \end{scope}
    \begin{scope}[shift={(1,0)}]
      \braiding
      \draw[thin](-2,0)--(-2,{sqrt(3)});
      \node[left] at (-2.2,{sqrt(3)/2}){\small $T_{i+1}$};  
    \end{scope}
    \nodes
  \end{scope}
  \end{tikzpicture}
  \end{center}
  \vspace{-10pt}
  \caption{Illustration of the defining algebra of the braid group $B_{N}$: $T_iT_{i+1}T_i = T_{i+1}T_iT_{i+1}$.}
  \label{fig:pfBraidingDef}
\end{figure}

For fermions, $\theta=\pi$, and we assign a minus sign to each interchange.  For bosons, $\theta=0$, and we assign no phase.  For all other values of the statistical parameter $\theta$, we assign a fractional phase factor $e^{\ii\theta}$ for each counterclockwise interchange of consecutively indexed particles in the relative amplitude in the many particle path integral.  We say the particles obey fractional statistics, and call them anyons.  Non-Abelian anyons, which we will discuss below, realize higher dimensional representations of the braid group $B_N$.

The most direct physical manifestation of the fractional statistics is the quantization of the kinetic (or dynamic) relative angular momentum of the anyons.  In 3D, the relative angular momentum is quantized as $\hbar l$, where $l$ is an odd integer for fermions, and an even integer for bosons.  In 2D, the amplitude acquires a statistical phase $e^{\ii\theta\varphi/{\pi}}$ as two anyons wind counterclockwise around each other with winding angle $\varphi$ in the xy-plane, and with the canonical relative angular momentum given by $L_\z=-\ii\hbar\partial_\varphi$, the kinetic relative angular momentum is quantized as\footnote{For a derivation of the sign of the second term, see Eqs.\ \eqref{eq:a_tube} to \eqref{eq:L_tube} below.}
\begin{align}
  \label{eq:lzrel}
  l_\z  = \hbar\left(\text{even integer}-\frac{\theta}{\pi}\right).
\end{align}
In other words, the relative angular momenta assumes odd multiples of $\hbar$ for fermions, even multiples of $\hbar$ for bosons, and fractionally shifted values for anyons.


Note that the rigorous possibility of fractional statistics exists only for quantum particles whose motion is essentially two dimensional.   This typically arises when motion in the third dimension is separated by a gap in the energy spectrum, and the relevant energies are below that gap.  An example of strictly two dimensional ``particles'' are vortices in an (approximately) two-dimensional quantum fluid.

\section{Charge-flux tube composites}

So far, our discussion of anyons was rather academic.  We have shown that the (path integral) formalism of quantum theory does allow for the possibility of fractional statistics, but have not given any context how nature would realize anyons.  When one of us\cite{wilczek82prl1144,wilczek82prl957} unknowingly rediscovered the possibility of fractional statistics in 1982, it was in the context of composites of charged particles with infinitesimally thin magnetic flux tubes.  These do not only provide a formal realization of anyons in 2D, but epitomize them.

To see this, consider first an otherwise free particle of charge $q$ confined to the xy-plane pierced by a magnetic flux tube of strength $\Phi=\frac{\theta}{\pi}\Phi_0$, where $\Phi_0=\frac{2\pi\hbar c}{q}$ is the Dirac quantum, at the origin.  In symmetric gauge, the vector potential is given by 
\begin{align}
  \label{eq:a_tube}
  \bs{a}(\bs{r}\!)=\frac{\Phi}{2\pi}\frac{\bs{e}_\z\times\bs{r}}{r^2}
  =\frac{\Phi}{2\pi r}\,\bs{e}_\varphi.
\end{align}
The magnetic field 
$\bs{b}(\bs{r}\!)=\nabla\times \bs{a}=\Phi\delta(\bs{r}\!)\,\bs{e}_\z$ 
is confined to the interior of the infinitesimal flux tube.  The Hamiltonian of the charged particle is
\begin{align}
  \label{eq:H_tube}
  H=\frac{1}{2m}\left(\bs{p}-\frac{q}{c}\bs{a}(\bs{r}\!)\right)^2
  =-\frac{\hbar^2}{2m}\frac{1}{r}\partial_r r\partial_r 
  + \frac{1}{2mr^2} L_z^2(\theta),
\end{align}
where
\begin{align}
  \label{eq:L_tube}
  L_z(\theta)=e^{+\ii\theta\varphi/{\pi}} 
  \left(-\ii\hbar\partial_\varphi \right) e^{-\ii\theta\varphi/{\pi}}
  =\hbar\left(-\ii\partial_\varphi-\frac{\theta}{\pi}\right)
\end{align}
is the kinetic angular momentum around the origin.  We see that the flux tube induces a fractional shift in accordance with \eqref{eq:lzrel}.
When we move the charge $q$ counterclockwise around the flux tube, the wave function picks up an Aharonov--Bohm phase 
\begin{align}
  \label{eq:flux_AB_phase}
  \frac{q}{\hbar c}\oint\bs{a}(\bs{r}\!)d\bs{r}
  = \frac{q\Phi}{\hbar c}=2\theta,
\end{align}
as required by \eqref{eq:tau1D} for a full counterclockwise winding of one anyon with statistical para\-meter $\theta$ around another.

This simple exercise suggests that we can transmute the statistics of a many particle system in 2D by attaching both charges $q$ and flux tubes of strength $\Phi=\frac{\theta}{\pi}\Phi_0$ 
to the particles.  Formally, we introduce a fictitious vector potential
\begin{align}
  \label{eq:a_fict}
  \bs{a}(\bs{r_i})
  =\frac{\theta}{\pi}\frac{\hbar c}{q} \sum_{\substack{ j\\(j\ne i)}}
  \frac{\bs{e}_\z\times(\bs{r}_i-\bs{r}_j)}{|\bs{r}_i-\bs{r}_j|^2},
\end{align}
which effects that particle $i$ sees fictitious flux tubes at the positions of all other particles $j$.  The many anyon Hamiltonian is then given by 
\begin{align}
  \label{eq:H_fict}
  H=\frac{1}{2m}\sum_i \left(\bs{p_i}-\frac{q}{c}\bs{a}(\bs{r}_i)
  +\frac{e}{c}\bs{A}(\bs{r}_i)\right)^2,
\end{align}
where we have included a coupling to an external, electromagnetic vector potential $\bs{A}(\bs{r})$, with electron charge $-e$. 

Strictly speaking, as we interchange two anyons consisting of charge-flux tube composites, we obtain a phase for each charge as it moves in the vector potential of the other flux tube.  The usual view assumed when we consider the model described by \eqref{eq:H_fict} with \eqref{eq:a_fict}, however, is that each particle moves in the vector potential of all the other flux tubes, while we 
ignore
the effect of the flux attached to the moving particle on all the other particles.  
The profound reason is that when we attach the fluxes to the charges in Chern--Simons theory, we find winding phases and kinetic relative angular momenta according to this ``simple'' counting, due to a field correction to the induced statistics\cite{goldhaber-89mpla21}.
On a more mundane level, this counting is consistent with both the statistics we obtain for the quasiparticles in the fractional quantum Hall effect\cite{halperin84prl1583,arovas-84prl722} (see \eqref{eq:qh_lzrel} below), and an exact model of charge-flux tube composites which connects integer and fractional quantum Hall states through an adiabatic process of attaching flux tubes to the charges\cite{greiter-21prbL121111}.

The connection between anyons and charge-flux tube composites becomes even more apparent on closed surfaces, such as spheres and tori\cite{einarsson90prl1995,greiter-21prbL121111}.  There, the allowed values for the number of anyons $N$ and the statistical parameter $\theta$ are restricted such that the \emph{total} flux---that is the sum of the fictitious and the electromagnetic flux---through the surface seen by each particle is a multiple of the Dirac flux quantum $\Phi_0$.  
%
%

\section{Chern--Simons construction and field correction}

A very elegant way to implement the previous construction in a local field theory is through a Chern--Simons (CS) term in a local field theory\cite{arovas-85npb117,goldhaber-89mpla21,dunne99proc}.  Consider a theory in 2+1 dimensions with a U(1) conserved particle current $J^\mu =(\rho,\bs{J})$, where $\rho$ is the particle density and $\bs{J}$ the particle current.  Now introduce a CS term in a fictitious gauge field $a^\mu (a^0,\bs{a})$ 
coupled with charge $q$ to this current by adding the following terms to the Lagrangian density,
\begin{align}
  \label{eq:CS_Delta_Lag}
  \Delta\mathcal{L}=-\frac{q}{c} J^\mu a_\mu 
  + \frac{\mu}{2c}\epsilon^{\mu\nu\rho}a_\mu\del_\nu a_\rho.
\end{align}
Then the Euler--Lagrange equation of motion for the field $a^\mu$ is
\begin{align}
  \label{eq:CS_Euler_Lag}
  q J^\mu = \mu \epsilon^{\mu\nu\rho}\del_\nu a_\rho.
\end{align}
The $\mu=0$ component 
couples the fictitious magnetic field to the charge,
\begin{align}
  \label{eq:CS_Euler_Lag_0}
  b_\z(\bs{r}) = \del_x a^y - \del_y a^x = -\frac{q}{\mu}\,\rho(\bs{r}),
\end{align}
and hence attaches a flux tube with $\Phi=-\frac{q}{\mu}$ to each point particle.  In other words, adding \eqref{eq:CS_Delta_Lag} to the Lagrangian turns the particles into charge-flux tube composites.

Let us now evaluate the phase generated by $\Delta\mathcal{L}$ in the path integral as we adiabatically wind one particle counterclockwise around another.  By definition, we obtain a phase factor
\begin{align}
  \label{eq:CS_e^iS}
  \exp\left({\frac{\ii}{\hbar}\Delta S}\right)=
  \exp\left(
  {\frac{\ii}{\hbar}\int\!\d t\d\bs{r}\Delta\mathcal{L}}\right).
\end{align}
We again use a symmetrical gauge, 
$a^0=0$, $\bs{a}=\frac{\Phi}{2\pi r}\,\bs{e}_\varphi$.
With
$\bs{J}(\bs{r}\!)=\rho\bs{v}=\rho r\del_t\varphi\,\bs{e}_\varphi$,
the first term in \eqref{eq:CS_Delta_Lag} yields the familiar Aharonov--Bohm phase for the motion of each charge in the vector potential of the other flux tube, 
\begin{align}
  \label{eq:CS_AB_phase_single}
  \frac{q}{\hbar c}\frac{\Phi}{2\pi }\int\!\d\bs{r} \rho 
  \int\!\d t\, \del_t\varphi
  =\frac{q}{\hbar c}\frac{\Phi}{2\pi }\varphi.
\end{align}
For an interchange with winding angle $\varphi=\pi$, counting the phases from both charges, we obtain a total phase factor
\begin{align}
  \label{eq:CS_AB_phase_total}
  \exp\left(\frac{\ii q\Phi}{\hbar c}\right)
  =\exp\left(2\ii\theta\right).
\end{align}
A more subtle contribution comes from the second term in \eqref{eq:CS_Delta_Lag}, the CS term.  Substituting \eqref{eq:CS_Euler_Lag} and evaluating it along the same lines, we see immediately that its contribution is minus one half of the Aharonov--Bohm phase \eqref{eq:CS_AB_phase_total}.  So the total phase we obtain when we adiabatically interchange two anyons via counterclockwise winding is $\exp\left(\ii\theta\right)$, and hence exactly what we obtain when we just count the phase of one particle in the fictitious vector potential of the other in \eqref{eq:H_fict} with \eqref{eq:a_fict}.

There is a simple physical reason for the field correction obtained here\cite{goldhaber-89mpla21}.  Recall that the flux induced by the CS term is proportional to the charge, $\Phi=-\frac{q}{\mu}$.  Now consider attaching the fictitious flux to it, and hence also the charge, adiabatically.  Each flux tube will generate a fictitious electric field
\begin{align}
  \oint \bs{E}\,\mathrm{d}\bs{s}=E_{\varphi}\cdot 2\pi r
  =-\frac{1}{c}\parder{\Phi}{t},
\end{align}
which in turn will change the kinematic angular momentum $l_\z $ by
\begin{align}
  \label{eq:CS_Delta_lz}
  \Delta l_\z 
  =\int\! F_{\varphi}r\,\d t
  =-\frac{1}{2\pi c}\int\! q(\Phi)\parder{\Phi}{t}\,\d t 
  =\frac{\mu}{2\pi c}\int\!\Phi\,\d\Phi 
  =-\frac{1}{2\pi c}\frac{q\Phi}{2}
  =-\frac{\hbar}{2}\frac{\theta}{\pi}, 
\end{align}
where we have substituted $\Phi=\frac{\theta}{\pi}\Phi_0$ in the last step.  As compared to \eqref{eq:L_tube}, the shift induced by the flux is halved.  The reason is that during the process, the charge is proportional to the flux, and even for $\del_t\Phi=\text{const.}$, the torque applied increases linearly with the flux.  Taking the effect of both fluxes on both charges into account, we multiply \eqref{eq:CS_Delta_lz} by 2, and recover \eqref{eq:L_tube}.

\section{Abelian anyons in fractionally quantized Hall states}

While Abelian and non-Abelian anyons also appear in models of 2D spin liquids\cite{kalmeyer-87prl2095,kitaev06ap2,greiter-09prl207203}, the most fully realized examples are the fractionally charged quasiparticle excitations of fractionally quantized Hall states.  In these systems, anyons have been observed in recent, groundbreaking experiments, as we will detail in the penultimate section.  Here, we give a brief account of the fractional statistics\cite{halperin84prl1583,arovas-84prl722,Laughlin90bookWilczek} of the quasiholes of Laughlin states\cite{laughlin83prl1395,greiter23proc_laughlin}, which describe quantized Hall liquids at Landau level filling fractions $\nu=1/m$, where $m$ is an odd integer.  We begin with a brief introduction to Landau levels (LLs) and the Laughlin wave function.    

\subsection*{Landau levels}

Consider an electron of charge $-e$ and mass $M$ confined to the xy-plane, and subject to a homogenous, perpendicular magnetic field $\bs{B}=-B\bs{e}_\z$.  It is convenient to introduce the complex coordinates $z=x+\ii y$ and $\bar z=x-\ii y$, their associated derivative operators 
$\del=\frac{1}{2}\left(\partial_x-\ii\partial_y\right)$, $\bdel=\frac{1}{2}\left(\partial_x+\ii\partial_y\right)$,
and the magnetic length $l=\sqrt{\frac{\hbar c}{eB}}$.  
In symmetric gauge, we may express the kinetic Hamiltonian in terms of ladder operators, \begin{align}
  \label{eq:LL_single_particle}
  H=\frac{1}{2M}\left(\bs{p}+\frac{e}{c}\bs{A}(\bs{r}\!)\right)^2
  =\hbar \omega_{\mathrm{c}} \left(a^\dagger a+\frac{1}{2}\right),
\end{align}
where $\omega_{\mathrm{c}}=\frac{eB}{Mc}$ is the cyclotron frequency and the ladder operators
\begin{align}
  \label{eq:LL_a_ladder}
  a=\frac{l}{\sqrt{2}}\left(2\bdel+\frac{z}{2l^2}\right), \quad
  a^\dagger=\frac{l}{\sqrt{2}}\left(-2\del+\frac{1}{2l^2}\bar z\right),  
\end{align}
obey $\comm{a}{a^\dagger}=1$.  It is readily seen that a complete (but unnormalized) basis of the eigenstates in the lowest Landau level (LLL, \ie with energy $\half\hbar\omega_{\mathrm{c}}$) is given by 
\begin{align}
  \label{eq:LL_basis}
  \psi_m(z)=z^m\,e^{-\frac{1}{4l^2}|z|^2},
\end{align}
where $n$ is a non-negative integer.  These states describe narrow rings 
centered around the origin, with radius $r_m=\sqrt{2m}\,l$.  
The areal density of states in each LL is given by 
\begin{align}
  \label{eq:LL_areal_density}
  \frac{\text{number of states}}{\text{area}}
  =\frac{m}{\pi r_m^2}=\frac{1}{2\pi l^2},
\end{align}
and the magnetic flux required for each state, $2\pi l^2 B =\frac{2\pi \hbar c}{e} =\Phi_0$, is hence given by the Dirac flux quantum.  In the following, we set the magnetic length $l=1$.  The wave function for a circular droplet of $N$ electrons in the LLL is given by
\begin{align}
  \label{eq:LL_filled}
  \psi(z_1,\hdots,z_N)
   =\mathcal{A}\left\{z_1^0z_2^1\ldots z_N^{N-1}\right\}\cdot
   \prod_{i=1}^N e^{-\frac{1}{4}|z_i|^2}
   =\prod^N_{i<j}(z_i-z_j)\,
   \prod_{i=1}^N e^{-\frac{1}{4}|z_i|^2},
\end{align}
where $\mathcal{A}$ denotes antisymmetrization.

The most general $N$ particle state in the LLL is given by 
\begin{align}
  \label{eq:LL_N_particle}
  \psi(z_1,\hdots,z_N)=f(z_1,\hdots,z_N) \prod_{i=1}^N e^{-\frac{1}{4}|z_i|^2},
\end{align}
where $f(z_1,\hdots,z_N)$ is analytic in all the $z$'s, and symmetric or antisymmetric for bosons or fermions, respectively.  If we impose periodic boundary conditions\cite{haldane-85prb2529}, we find that\linebreak $\psi(z_1,z_2,\hdots,z_N)$, when viewed as a function of $z_1$ while $z_2,\hdots,z_N$ are parameters, has exactly as many zeros as there are states in the LLL, \ie as there are Dirac flux quanta going through the principal region.
If $\psi(z_1,\hdots,z_N)$ describes fermions and is hence antisymmetric, there will be at least one zero seen by $z_1$ at each of the other particle positions.  The most general wave function is hence
\begin{align}
  \label{eq:LL_N_fermions}
  \psi(z_1,\hdots,z_N)
  =P(z_1,\hdots,z_N) \prod^N_{i<j}(z_i-z_j) \prod_{i=1}^N e^{-\frac{1}{4}|z_i|^2},
\end{align}
where $P$ is a symmetric polynomial in the $z_i$'s.  In the case of 
a completely filled Landau level, there are only as many zeros as there
are particles, which implies that all except one of the zeros in $z_1$ 
will be located at the other particle positions  $z_2,\hdots,z_N$.
This yields \eqref{eq:LL_filled} as the unique state for open
boundary conditions.

\subsection*{The Laughlin wave function}

The experimental observation\cite{tsui-82prl1559,papic-23proc} which Laughlin's theory\cite{laughlin83prl1395,greiter23proc_laughlin} explains is a plateau in the Hall resistivity of a two-dimensional electron gas at 
$\rho_{\x\y}={3h}/{e^2}$, 
\ie at a Landau level filling fraction $\nu=1/3$.  The filling fraction denotes the number of particles divided by the number of number of states in each Landau level in the thermodynamic limit, and is defined through 
\begin{align}
  \label{eq:qhnu}
  \frac{1}{\nu}=\parder{N_\Phi}{N},
\end{align}
where $N_\Phi$ is the number of Dirac flux quanta through the sample and $N$ is the number of particles.  For a wave function at $\nu=1/3$, we consequently have three times as many zeros seen by $z_1$ as there are particles, and the polynomial $P(z_1,\hdots,z_N)$ in \eqref{eq:LL_N_fermions} has two zeros per particle.  The experimental findings, as well as early numerical work by Yoshioka, Halperin, and Lee\cite{yoshioka-83prl1219}, are consistent with, if not indicative of, a quantum liquid state at a preferred filling fraction $\nu=1/3$.  Since the kinetic energy is degenerate in each Landau level, such a liquid has to be stabilized by the repulsive Coulomb interactions between the electrons.  This implies that the wave function should be highly effective in suppressing configurations in which particles approach each other, as there is a significant potential energy cost associated with it.  We may hence ask ourselves whether there is any particular way of efficiently distributing the zeros of $P(z_1,\hdots,z_N)$ in this regard.

Laughlin's wave function amounts to attaching the additional zeros onto the particles, such that each particle coordinate $z_2,\hdots,z_N$ becomes a triple zero of $z_1$ when $\psi(z_1\hdots,z_N)$ is viewed as a function of $z_1$ with parameter $z_2,\hdots,z_N$.  For filling fraction $\nu=1/m$, where $m$ is an odd integer if the particles are fermions and an even integer if they are bosons, Laughlin proposed the ground state wave function 
\begin{align}
  \label{eq:psi_Laughlin}
  \psi_m(z_1,\ldots,z_N)
  =\prod^N_{i<j}(z_i-z_j)^m\prod_{i=1}^N e^{-\frac{1}{4}|z_i|^2}.
\end{align}
There are hence no zeros wasted---all of them contribute in keeping the particles away from each other effectively, as $\psi_m$ vanishes as the $m$-th power of the distance when two particles approach each other.  
This is the uniquely defining property of the Laughlin state, and also the property which enabled Haldane\cite{haldane83prl605} to identify a parent Hamiltonian, which singles out the state as its
unique and exact ground state.  
It describes an incompressible quantum liquid, as the construction is only possible at filling fractions $\nu=1/m$.

Even within the LLL limit, which we assume to hold in our discussion, the Laughlin state \eqref{eq:psi_Laughlin} is not the exact ground state for electrons with (screened) Coulomb interactions at filling fraction $\nu=1/3$.  It is, however, reasonably close in energy and has a significant overlap with the exact ground state for finite systems.  The difference between the exact ground state and Laughlin's state is that in the exact ground state, the zeros of $P(z_1,z_2,\hdots,z_N)$ are attached to the particle coordinates, but do not coincide with them\cite{greiter97pe1}.  At long distances, the physics described by both states is identical.  In particular, the topological quantum numbers of both states, such as the charge and the statistics of the (fractionally) charged excitations, or the degeneracies on closed surfaces of genus one and higher, are identical.

\subsection*{Fractionally charged quasiparticle excitations}

Laughlin\cite{laughlin83prl1395} created the elementary, charged excitations of the fractionally quantized Hall state \eqref{eq:psi_Laughlin} through a \emph{Gedankenexperiment}.  If one adiabatically inserts one Dirac quantum $\Phi_0=\frac{hc}{e}$ of magnetic flux in the $\bs{e}_\z$ direction through an infinitesimally thin solenoid at a position $\xi$, and then removes this flux quanta via a singular gauge transformation, the final Hamiltonian will be identical to the initial one.  The final state will hence be an eigenstate of the initial Hamiltonian as well.  The adiabatic insertion of the flux will induce an electric field 
\begin{align}
  \oint \bs{E}\,\d\bs{s}=E_{\varphi}\cdot 2\pi r
  =-\frac{1}{c}\parder{\Phi}{t}.
\end{align}
Since the electrons are confined to the LLL, this will not lead to an increase in the kinetic angular momentum as in \eqref{eq:CS_Delta_lz},
but to a perpendicular current
$J_{\text{r}}=\sigma_{\x\y}E_\varphi$, where $\sigma_{\x\y}=\frac{1}{m}\frac{e^2}{h}$.\footnote{Since the magnetic field 
is oriented in the $-\bs{e}_\z$ direction, $\rho_{\x\y}<0$ and $\sigma_{\x\y}>0$.}
The charge transported away from the center of the flux tube is
\begin{align}
  \label{eq:QH_Q_transported}
  \Delta Q=2\pi r\int\! J_{\text{r}}\,\d t
  =-\frac{1}{m}\frac{e^2}{hc}\int
  \! \d\Phi = -\frac{e}{m}
\end{align}
Since the Landau level filling fraction is $\nu=1/m$, a fractional charge of $-e/m$ occupies one state in the LLL.
If we choose a basis of eigenstates of angular momentum around $\xi$, the basis states evolve according to 
\begin{align}
  \label{eq:qhLzBasisEvo}
  (z-\xi)^m \,e^{-\frac{1}{4}|z|^2}
  \rightarrow \abs{z-\xi}\cdot (z-\xi)^m \,e^{-\frac{1}{4}|z|^2}
  \rightarrow (z-\xi)^{m+1} \,e^{-\frac{1}{4}|z|^2},
\end{align}
where the last step is due to the singular gauge transformation.  The process increases the canonical relative angular momentum around $\xi$ by $\hbar$.


The Laughlin ground state \eqref{eq:psi_Laughlin} evolves in the process into 
\begin{align}
  \label{eq:qhpsiQH}
  \psi_\xi^{\s\text{QH}}(z_1,\hdots,z_N)
  =\prod^N_{i=1}(z_i-\xi)
  \prod^N_{i<j}(z_i-z_j)^m\prod_{i=1}^N e^{-\frac{1}{4}|z_i|^2},
\end{align}
which describes a quasihole excitation at $\xi$.  It is easy to confirm
that if the electron charge is $-e$, the charge of the quasihole is
$+e/m$, as obtained in \eqref{eq:QH_Q_transported}.  If we were to create $m$ quasiholes at $\xi$ by inserting $m$ Dirac quanta, the final wave function would be
\begin{align}
  \label{eq:qhpsiQHThree}
  \psi_\xi^{\s m\,\text{QH's}}(z_1,\hdots,z_N)
  =\prod^N_{i=1}(z_i-\xi)^m
  \prod^N_{i<j}(z_i-z_j)^m\prod_{i=1}^N e^{-\frac{1}{4}|z_i|^2},
\end{align}
\ie we would have created a true hole in the liquid, which is screened
as all the other electrons.  Since the hole has charge $+e$,
the quasihole has charge $+e/m$.  One may view the quasihole as a zero
in the wave function which is not attached to any of the electrons.

The quasielectron, \ie the antiparticle of the quasihole, has charge
$-e/m$ and is created by inserting the flux adiabatically in the
opposite direction, thus lowering the angular momentum around some
position $\xi$ by $\hbar$, or alternatively, by removing one of the
zeros from the wave function.  For ease in presentation, we will limit our discussion here to quasihole excitations.

\subsection*{Fractional statistics of quasihole excitations}

When Laughlin introduced the quasiparticle excitations of quantized Hall states, he introduced them as localized defects or more precisely, vortices in an otherwise uniform quantum liquid.  To address the question of their statistics, however, it is necessary to view them as particles, with a Hilbert space spanned by the parent wave function for the electrons.  We consider here a Laughlin state with two quasiholes in an eigenstate of relative angular momentum in an ``orbit'' centered at the origin.  Since the quasiholes have charge $e^*=+e/m$, the effective flux quantum seen by them is $\Phi_0^*=\frac{2\pi\hbar c}{e^*}=m\Phi_0$,
and the effective magnetic length is $l^*=\sqrt{\frac{\hbar c}{e^*B}} =l\sqrt{m}$. 
We expect the single quasihole wave function to describe a particle of
charge $e^*$ in the LLL, and hence be of the general form
\begin{align}
  \label{eq:qhQHphi}
  \phi(\bar\xi)=f(\bar\xi)\,e^{-\frac{1}{4m}|\xi|^2}.
\end{align}
The complex conjugation 
reflects that the sign of the quasihole charge is reversed relative to
the electron charge $-e$.  

The electron wave function for the state with two quasiholes 
in an eigenstate of relative angular momentum is given by
\begin{align}
  \label{eq:qh2QH}
  \psi(z_1,\hdots,z_N)
  =\int \text{D}[\xi_1,\xi_2] 
  \,\phi_{p,m}(\bar\xi_1,\bar\xi_2)
  \,\psi_{\xi_1,\xi_2}^{\s\text{QHs}}(z_1,\hdots,z_N)
\end{align}
with
\renewcommand{\strut}{\rule[-.6\baselineskip]{0pt}{2\baselineskip}}
\begin{align}
  \label{eq:qhphipm}\strut
  \phi_{p,m}(\bar\xi_1,\bar\xi_2)
  =(\bar\xi_1-\bar\xi_2)^{p+\frac{1}{m}} \prod_{k=1,2}e^{-\frac{1}{4m}|\xi_k|^2},
\end{align}
where $p$ is an even integer, and
\begin{align}
  \label{eq:qhpsiXi1Xi2}
  \psi_{\xi_1,\xi_2}^{\s\text{QHs}}(z_1,\hdots,z_N)
  =(\xi_1-\xi_2)^{\frac{1}{m}}\prod_{k=1,2}\left(e^{-\frac{1}{4m}|\xi_k|^2}
  \prod^N_{i=1}(z_i-\xi_k)\right) 
  \prod^N_{i<j}(z_i-z_j)^m\prod_{i=1}^N e^{-\frac{1}{4}|z_i|^2}.
\end{align}
The quasihole coordinate integration extends over the complex plane,
\begin{equation*}
 \label{eq:qhD}
 \int \text{D}[\xi_1,\xi_2]
 \equiv \int\!\!\ldots\!\!\int 
 \text{d}x_1 \text{d}y_1 \text{d}x_2 \text{d}y_2, 
\end{equation*}
where $\xi_1=x_1+\ii y_1$ and $\xi_2=x_2+\ii y_2$.

This requires some explanation.  We see that both $\phi_{p,m}(\bar\xi_1,\bar\xi_2)$ and $\psi_{\xi_1,\xi_2}^{\s\text{QHs}}(z_1,\hdots,z_N)$ contain multiple valued functions of $\bar\xi_1-\bar\xi_2$ and $\xi_1-\xi_2$, respectively, while the product of them is understood to be single valued.  The reason is that the Hilbert space for the quasiholes at $\xi_1$ and $\xi_2$ spanned by $\psi_{\xi_1,\xi_2}^{\s\text{QHs}}(z_1,\hdots,z_N)$ has to be normalized and is, apart from the exponential, supposed to be analytic in $\xi_1$ and $\xi_2$.  At the same time, we expect $\phi_{p,m}(\bar\xi_1,\bar\xi_2)$ to be of the general form \eqref{eq:qhQHphi}, \ie to be an analytic function of $\bar\xi_1$, $\bar\xi_2$ times the exponential.

The form \eqref{eq:qhphipm} of the quasihole wave function including its branch cut, is indicative of fractional statistics with statistical parameter $\theta =\pi/m$, as the canonical relative angular momentum\footnote{Particles in the LLs are special in that their kinetic angular momentum is given by $\hbar$ times the LL index, and zero in the LLL.  Therefore, fractional statistics manifests itself only in the canonical relative angular momentum.} of the quasiholes is given by \begin{align}
  \label{eq:qh_lzrel}
  l_\z= -\hbar\left(p + \frac{1}{m}\right).
\end{align}
This result agrees with the results of Halperin\cite{halperin84prl1583} and of Arovas, Schrieffer, and one of us\cite{arovas-84prl722}, who calculated the statistical parameter directly using the adiabatic theorem\cite{berry84prsla45,ShapereWilczek89}.


Note that if we view the quasiholes as charge-flux tube composites with charge $e/m$ 
and flux $\Phi_0=\frac{2\pi\hbar c}{e}$, the value $\theta =\pi/m$ accounts only for the phase one charge would acquire in the presence of the other flux tube, without a factor of 2, and as obtained with the 
field correction in CS theory\cite{goldhaber-89mpla21}.  To see why this is correct and consistent, imagine the simultaneous creation of two quasiholes by the insertion of magnetic fluxes at positions $\xi_1$ and $\xi_2$.  During the process, both the charge of the quasiholes and the flux in the tubes will be ramped up simultaneously, from $0$ to $e/m$ 
and from $0$ to $\Phi_0$, respectively.  We hence have exactly the situation we had in CS-theory, and obtain a statistical phase in accordance with it.

\section{Anyons in one dimension}

Just as fractional statistics was considered impossible for more than half a century after the discovery of quantum mechanics, a generalization of the concept to one dimension has long seemed out of reach.  The reason is simply that while in 2D, different windings between particles are topologically distinct, there is no analog of a winding number when we exchange particles in one dimension (1D).  The only topologically distinct classes of paths, it seems, are those where we interchange the particles, and those where we do not interchange them.  Since interchanging them twice is equivalent to not interchanging them at all, we are left with the familiar cases of bosons and fermions.

There are various manifestations of the fractional phases the wave functions of anyons acquire.  The simplest is fractional relative angular momenta (see Equation \eqref{eq:lzrel}).  Another manifestation is a fractional exclusion principle for anyons, which for fermions is known as the Pauli principle.  
If the statistics is $\theta/\pi=p/q$, the creation of $q$ anyons will reduce the number of orbitals available for further anyons by $p$.  The fractional exclusion is so natural that it has hardly been mentioned in the literature on anyons in 2D.

In 1991, to the surprise of many, Haldane\cite{haldane91prl937} discovered that the spinon excitations of an exactly solvable model of a spin chain, the Haldane--Shastry model (HSM),\cite{haldane88prl635,shastry88prl639,haldane-92prl2021,Greiter11} exhibit a fractional exclusion principle.  If we embed a spin $\half$ chain with periodic boundary conditions and an even number of sites $N$ as a unit circle in the complex plane,
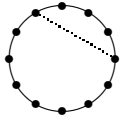
\begin{center}
  \begin{picture}(320,70)(-40,-35)
    \put(0,0){\circle{100}}
    \put( 20.0,   .0){\circle*{3}}
    \put( 17.3, 10.0){\circle*{3}}
    \put( 10.0, 17.3){\circle*{3}}
    \put(   .0, 20.0){\circle*{3}}
    \put(-10.0, 17.3){\circle*{3}}
    \put(-17.3, 10.0){\circle*{3}}
    \put(-20.0,   .0){\circle*{3}}
    \put(-17.3,-10.0){\circle*{3}}
    \put(-10.0,-17.3){\circle*{3}}
    \put(   .0,-20.0){\circle*{3}}
    \put( 10.0,-17.3){\circle*{3}}
    \put( 17.3,-10.0){\circle*{3}}
    \qbezier[20]( 20.0,   .0)(  5.0,8.65)(-10.0, 17.3)
    \put(50,12){\makebox(0,0)[l]
      {$N$\ sites with spin $\half$ on unit circle: 
      }}
    \put(50,-12){\makebox(0,0)[l]
      {$\displaystyle \eta_\alpha=e^{\text{i}\frac{2\pi}{N}\alpha }$
        \ \ with\ $\alpha = 1,\ldots ,N$}}
  \end{picture}
\end{center}
the Haldane--Shastry Hamiltonian is given by 
\begin{align}
  \label{eq:hsham}
  H=\left(\frac{2\pi}{N}\right)^2\sum^N_{\alpha <\beta}\,
  \frac{{\bs{S}}_\alpha {\bs{S}}_\beta}{|\eta_\alpha-\eta_\beta|^2}\,,
\end{align}
where $|\eta_\alpha-\eta_\beta|$ is the chord distance between
the sites $\alpha$ and $\beta$.  The ground state 
is given by
\begin{align}
  \label{eq:hspsi0}
  \psi_{0}(z_1,z_2,\ldots ,z_M) = 
  \prod_{i<i}^M\,(z_i-z_j)^2\,\prod_{i=1}^M\,z_i, 
\end{align}
where the $z_i$'s denote the $M=\frac{N}{2}$ $\up$ spin coordinates on the unit circle.  The wave function for a $\dw$ spin spinon localized at site  $\eta_\alpha$ is then given by
\begin{align}
  \label{eq:hsp1sp}
  \psi_{\alpha\dw}
  (z_1,z_2,\ldots ,z_M) =\prod_{i=1}^M\,(\eta_\alpha -z_i)\,
  \psi^{\s\text{HS}}_{0}(z_1,z_2,\ldots ,z_M),
\end{align}
where $M=\frac{N-1}{2}$ is the number of $\up$ or $\dw$ spins condensed in the uniform liquid.  Since ${S}^\z _{\text{tot}} \psi_{\alpha\dw} = -\frac{1}{2} \psi_{\alpha\dw}$ and ${S}^-_{\text{tot}} \psi_{\alpha\dw} = 0$, the spinon transforms as a spinor under spin rotations.  It hence carries the spin, but not the charge, of the underlying electrons\cite{greiter23proc_spincharge}.  Spinons are fractionalized excitations, and both conceptually and mathematically similar to the quasiparticle excitations in fractionally quantized Hall fluids.  The fractional quantum number is the spin, which is one-half in a Hilbert space built out of spin-flips, which carry spin one.  The localized spinon (\ref{eq:hsp1sp}) is not an eigenstate of the
Hamiltonian (\ref{eq:hsham}).  To obtain exact
eigenstates, we construct momentum eigenstates according to
\begin{align}
  \label{eq:hspsim}
  \psi_{m\dw}(z_1,z_2,\ldots ,z_M) =
  \sum_{\alpha=1}^{N} (\bar\eta_\alpha)^m\,
  \psi_{\alpha\dw}(z_1,z_2,\ldots ,z_M), 
\end{align}
where the integer $m$ corresponds to a momentum quantum number.  Since $\psi_{\alpha\dw} (z_1,z_2,\ldots ,z_M)$ contains
only powers $\eta_\alpha^0, \eta_\alpha^1,\ldots , \eta_\alpha^M$ and
\begin{align}
  \label{eq:hsdelta}
  \sum_{\alpha=1}^{N} \overline{\eta}_\alpha^m \eta_\alpha^n = \delta_{mn}
  \quad \text{mod}\ N,
\end{align}
$\psi_{m\dw} (z_1,z_2,\ldots ,z_M)$ will vanish unless $m=0,1,\ldots ,M$.  In the limit of large system sizes, there are only half as many spinon orbitals as there are sites.

If we consider a state with $L$ spinons, we can easily see from \eqref{eq:hspsim} and \eqref{eq:hsdelta} that the number of orbitals available for further spinons we may wish to create is $M+1$, where $M=\frac{N-L}{2}$ is the number of $\up$ or $\dw$ spins in the remaining uniform liquid.  (In this representation, the spinon wave functions are symmetric; two or more spinons can have the same value for $m$.)  In other words, the creation of {\em two} spinons reduces the number of available single spinon states by {\em one}.  They hence obey half-Fermi statistics in the sense of Haldane's exclusion principle.

The HSM is formally extremely appealing, as there is no interaction between the excitations, and all the eigenstates can be identified exactly.  In this sense, the model is as unique to the world of spin chains as the free Fermi sea is to electronic states in three dimensional metals.  The Yangian symmetry of the
model~\cite{haldane-92prl2021} implies significant
degeneracies in the spectrum and hence indicates integrability.  The
model is not integrable in the usual sense, however, as the method of
quantum inverse scattering~\cite{KorepinBogoliubovIzergin93} is not
applicable to models with longe-range interactions.  Talstra and
Haldane~\cite{talstra-95jpa2369} have nonetheless succeeded in
constructing an infinite set of mutually commuting integrals of motion, which provide the framework for the
model's integrability.  

\begin{figure}[tb] 
  \begin{center}
  \pgfmathsetmacro{\rad}{.5}
  \begin{tikzpicture}[>=latex,scale=.8]
    \begin{scope}[shift={(0,0)}]
      \draw[thin,->] (0,-.1) node[below]{$0$} to 
      (0,2.4) node[below left]{$\epsilon(p)$};
      \draw[thin,->] (0,0) to (7,0) node[below]{$p$};
      \draw [thick,domain=0:pi] plot (\x,{.8*\x*(pi-\x)});
      \draw [thick] (pi,-.1) node[below]{$\pi$} -- (pi,0.1);    
      \draw [thick] (2*pi,-.1) node[below]{$2\pi$} -- (2*pi,0.1);    
      \draw [blue,thin] (1,2) to (1,-3.6) node[below]{$p_1$};
      \draw [red,thin] (2,2) to (2,-3.6) node[below]{$p_2$};
    \end{scope}
    \begin{scope}[shift={(0,-3)}]
      \draw[thin,->] (0,-1.4) to 
      (0,1.8) node[below left] {$v_{\text{g}}(p)$};
      \draw[thin,->] (0,0) to (7,0) node[below]{$p$};
      \draw [thick,domain=0:pi] plot (\x,{.8*(pi/2-\x)});
      \draw [thin] (pi,-.8*pi/2) to (pi,-.6);
      \draw [thick] (pi,-.1) node[below]{$\pi$} -- (pi,0.1);    
      \draw [thick] (2*pi,-.1) node[below]{$2\pi$} -- (2*pi,0.1);    
    \end{scope}
  \end{tikzpicture}
  \end{center}
  \caption{Spinon dispersion $\epsilon(p)$ and group velocity $v_{\text{g}}(p)=\del_p \epsilon(p)$ in the SU(2) Haldane--Shasty model.  The interval of allowed spinon momenta is restricted to a part of the Brillouin zone, and the group velocity is a monotonic function of momentum.}
  \label{fig:1D_spectra} 
\end{figure}
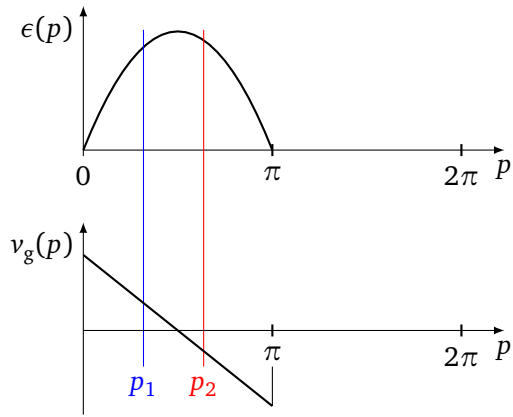 

For almost two decades, the consensus was that fractional statistics must exist in one dimension, but could only be defined through the generalized Pauli principle introduced by Haldane\cite{haldane91prl937}.  The connection to phases acquired by amplitudes in path integrals and topology was only understood by one of us\cite{greiter09prb064409} in 2009.  

\begin{figure}[tb] 
  \begin{center}
    \pgfmathsetmacro{\rad}{.5}
    \begin{tikzpicture}[>=latex,scale=.8]
      \begin{scope}[shift={(0,0)}]
      \draw[thin]( 0,0) circle[radius=1];
      \draw[fill,thick]( -0.866,-.5) circle[radius=.04];
      \draw[fill,thick]( 1,0) circle[radius=.04];
      \draw[blue,thick] (1.13,0) to [out=90,in=0] (0,1.13);
      \draw[blue,thick] (0,1.13) to [out=180,in=90] (-1.13,0);
      \draw[blue,thick,->] (-1.13,0) to [out=270,in=120](-.978,-.57);
      \draw[red,thick](-.745,-.43) to [out=300,in=180] (0,-.86); 
      \draw[red,thick,->](0,-.86) to [out=0,in=-92] (0.86,.04); 
      \node[right] at (1.15,0) {\small $1$};
      \node[left] at (-.98,-.68) {\small $2$};
      \node at (0,-1.7) {$1$};
      \end{scope}
      \begin{scope}[shift={(4,0)}]
      \draw [thin]( 0,0) circle[radius=1];
      \draw [fill,thick]( 0,1) circle[radius=.04];
      \draw [fill,thick]( 1,0) circle[radius=.04];
      \draw [blue,thick,->](1.14,0) to [out=90,in=358] (-.04,1.13);
      \draw [red,thick,->](0,.86) to [out=0,in=92] (0.86,-.04);
      \node[right] at (1.15,0) {\small $1$};
      \node[above] at (0,1.2) {\small $2$};
      \node at (0.2,-1.7) {$e^{\ii\theta}$};
      \end{scope}
      \begin{scope}[shift={(8,0)}]
      \draw [thin]( 0,0) circle[radius=1];
      \draw [fill,thick]( 0,1) circle[radius=.04];
      \draw [fill,thick]( 1,0) circle[radius=.04];
      \draw [blue,thick] (1.13,0) to [out=90,in=0] (0,1.13);
      \draw [blue,thick] (0,1.13) to [out=180,in=90] (-1.13,0);
      \draw [blue,thick] (-1.13,0) to [out=270,in=180] (.07,-1.20);
      \draw [blue,thick] (.07,-1.20) to [out=0,in=270] (1.26,0);
      \draw [blue,thick,->](1.26,0) to [out=90,in=358] (-.04,1.25);
      \draw [red,thick,->](0,.86) to [out=0,in=92] (0.86,-.04);
      \node[right] at (1.15,0) {\small $1$};
      \node[above] at (0,1.2) {\small $2$};
      \node at (0.2,-1.7) {$e^{2\ii\theta}$};
      \end{scope}
      \begin{scope}[shift={(12,0)}]
      \draw [thin]( 0,0) circle[radius=1];
      \draw [fill,thick]( 0,1) circle[radius=.04];
      \draw [fill,thick]( 1,0) circle[radius=.04];
      \draw [blue,thick](1.14,0) to [out=90,in=0] (0,1.14);
      \draw [blue,thick](0,1.14) to [out=180,in=90] (-1.14,0);
      \draw [blue,thick](-1.14,0) to [out=270,in=180] (0,-1.14);
      \draw[blue,thick,->] (0,-1.14) to [out=0,in=262] (1.138,-.02);
      \node[right] at (1.15,0) {\small $1$};
      \node[above] at (0,1.2) {\small $2$};
      \node at (0.2,-1.7) {$e^{\ii\theta}$};
      \end{scope}
    \end{tikzpicture}
  \end{center}
  \caption{Examples of topologically distinct windings of two particles in 1D for a concave spectrum with $\theta\in[0,\pi]$, $p_2>p_1$, and $v_{\text{g}}(p_2)<v_{\text{g}}(p_1)$, with and without interchanges, and the associated phase factors in accordance with \eqref{eq:1D_p_rel}.}
  \label{fig:1D_windings} 
\end{figure}
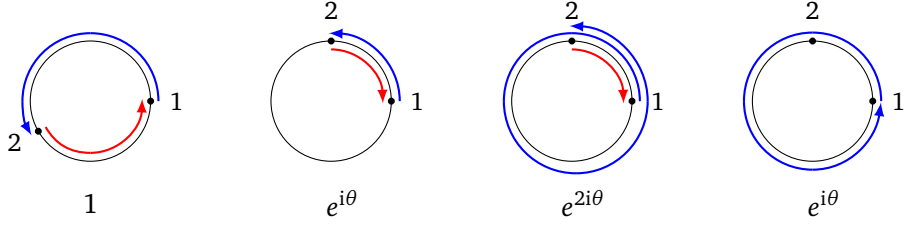

The HSM, and in fact all one dimensional models with anyon excitations, have very special spectra 
(see Figure \ref{fig:1D_spectra}).  Not all the single-particle momenta in the Brillouin zone for periodic boundary conditions are available.  In the HSM, for example, the allowed single spinon momenta span to only half of the Brillouin zone.  The single-particle spectrum in these models is always either entirely concave or entirely convex.  In the space of available momenta, the group velocity---that is, the velocity a wave packet build out of these excitations would have---is therefore always either a monotonically decreasing or a monotonically increasing function of single-particle momentum.  When we view velocities in clockwise direction as positive, one of the spinons is hence always faster than the other one, and the crossings are always uni-directional.  In the context of this uni-directionality, paths with different numbers of uni-directional crossings become topologically distinct, and we can assign a statistical phase $\theta$ to the crossings of anyons in 1D (see Figure \ref{fig:1D_windings}).  In analogy to the fractional shifts in the kinetic relative angular momenta in 2D (see Equation \eqref{eq:lzrel}), the spacings of the kinetic linear momenta on the circle are shifted by the fractional phases $\theta$ acquired as the anyons cross each other, 
\begin{align}
 \label{eq:1D_p_rel}
 \Delta p = p_2-p_1=\frac{2\pi\hbar}{L}
 \left(\text{non-negative integer}+\frac{\abs{\theta}}{\pi}\right),
\end{align}
where $p_2>p_1$, and $L$ is the circumference of the circle.  The sign of $\theta\in[\pi,\pi]$ is positive for concave spectra, where the group velocity $v_{\text{g}}(p)=\del_p \epsilon(p)$ decreases monotonically with $p$, and negative for convex spectra, where $v_{\text{g}}(p)$ increases monotonically with $p$.  For a system with only two anyons with $p_2>p_1$, the shifts induced by the fractional statistics for the individual momenta are 
\begin{align}
  \label{eq:1D_shifts_p1_p2}
  p_2\to p_2+\frac{2\pi\hbar}{L}\frac{\abs{\theta}}{2\pi},\quad
  p_1\to p_1-\frac{2\pi\hbar}{L}\frac{\abs{\theta}}{2\pi}.
\end{align}
For fermions with $\theta=\pi$, the creation of the second fermion on the ring changes the boundary conditions for the first from periodic to anti-periodic and vice versa.

Note that even though anyons in 1D are identical particles, they become dynamically distinguishable, that is, distinguishable through the quantum states they occupy.  This is the cost in conceptual complexity required for fractional statistics in 1D.

The general principles of fractional statistics in 1D outlined here\cite{greiter09prb064409} have only been worked out for spinons in the HSM\cite{bernevig-01prl3392,greiter-05prb224424} (where the spectrum is concave), and for holons in the Kuramota-Yokoyama model\cite{kuramoto-91prl1338,thomale-06prb024423} (KYM, where the spectrum is convex).  In these models, both the single anyon momenta $p_i$ and the fractional shifts $\theta/\pi$ are good quantum numbers.  In models with 1D anyons in general, which we may view as HSMs or KYMs with additional interactions between the excitations, we do not expect the single momenta $p_i$ to be good quantum numbers.  The fractional shifts in the momentum spacings \eqref{eq:1D_p_rel}, however, are topological invariants, which will not be affected by perturbations, and hence good quantum numbers.  Similar considerations apply to 2D, where the relative angular momenta of quasiholes in a fractionally quantized Hall state are in general not good quantum numbers, while the fractional shifts in \eqref{eq:lzrel} or \eqref{eq:qh_lzrel} are topological invariants.

\section{Non-Abelian anyons}

Non-Abelian anyons transform according to higher dimensional representations of the braid group \eqref{eq:pfBraidAlgebra} under braiding operations.  This is possible because the anyons span an internal Hilbert space.  To see how this works, we consider here the simplest case of non-Abelian anyons, the Ising anyons realized by the vortices of p-wave superfluids, and equivalently also by the quasiparticle excitations of the Pfaffian state in the quantum Hall effect.

\subsection*{The Pfaffian state}

The Pfaffian state at even-denominator Landau level filling fractions
was introduced independently by Moore and Read\cite{moore-91npb362}
as an example of a quantized Hall state which supports non-Abelian excitations, and by Wen and us\cite{greiter-91prl3205,greiter-92npb567} as a candidate for the observed plateau\cite{willett-87prl1776} in Hall resistivity at Landau level filling fraction $\nu=5/2$, \ie at $\nu=1/2$ in the second Landau level.

The wave function first proposed by Moore and Read\cite{moore-91npb362} is
\begin{align}
  \label{eq:pfpsi0}
  \psi^{\text{Pf}}(z_1,z_2,\ldots ,z_N) 
  =\text{Pf}\left(\frac{1}{z_{i}-z_{j}}\right)\,
  \prod_{i<j}^{N}(z_i-z_j)^2\,
  \prod_{i=1}^N e^{-\frac{1}{4}|z_i|^2},          
\end{align}
where the particle number $N$ is even, and the Pfaffian is is given by the fully antisymmetrized sum over all possible pairings of the $N$ particle coordinates
\begin{align}
  \label{eq:pfpfaff}
  \text{Pf}\left(\frac{1}{z_i -z_j}\right)\equiv
  \mathcal{A}
  \left\{
    \frac{1}{z_1-z_2}\cdot\,\ldots\,\cdot\frac{1}{z_{N-1}-z_{N}}
  \right\}.
\end{align}
The state describes a bosonic Laughlin state at $\nu=1/2$ supplemented by a Pfaffian, which implements p-wave pairing correlations.  Since the Pfaffian is completely antisymmetric, it reverses the statistics from bosons to fermions, but does not change the Landau level filling fraction.
More generally, a Pfaffian wave function $\psi(\bs{x}_1\ldots\bs{x}_{N})=\text{Pf}\left(\varphi(\bs{x}_i-\bs{x}_j)\right)$ describes a BCS wave function\cite{Schrieffer64} in position space, obtained from the usual BCS state by projecting out a definite number of particles\cite{DysonQuotedinSchrieffer64,greiter05ap217}.

\subsection*{Quasiparticle excitations and the internal Hilbert space}

One of the key properties of superconductors is that the magnetic vortices are quantized in units of one half of the Dirac flux quanta $\Phi_0=2\pi\hbar c/e$, in accordance with the charge $-2e$ of the Cooper pairs.  The pairing correlations in the Pfaffian Hall state have a similar effect on the vortices or quasiparticle excitations, which carry one half of the flux and charge they would carry without the pairing, \ie they carry charge $e^*=e/4$.  The wave function for two flux~$\half$ quasiholes at positions $\xi_1$ and $\xi_2$ is easily formulated.  We simply replace each factor in the Pfaffian in \eqref{eq:pfpsi0} by \begin{align}
  \label{eq:pf2QH}
  \text{Pf}\left(\frac{1}{z_i-z_j}\right) 
  \mapsto \text{Pf}\left(
    \frac{(z_i-\xi_1)(z_j-\xi_2)+(z_i\leftrightarrow z_j)}{z_i-z_j}
  \right),
\end{align}
such that one member of each electron pair sees the additionally inserted zero at $\xi_1$ and the other member sees it at $\xi_2$.  If we set $\xi_1=\xi_2=\xi$, we will recover a regular quasihole in the Laughlin fluid with charge $e^*=e/2$.

The internal Hilbert space spanned by the quasiparticle excitations only emerges as we consider the wave function for four charge $e^*=e/4$ quasiholes at positions $\xi_1,\ldots,\xi_4$, which is obtained by replacing the Pfaffian in \eqref{eq:pfpsi0} by \begin{align}
  \label{eq:pf4QH}
  \text{Pf}\left(\frac{1}{z_i-z_j}\right) 
  \mapsto \text{Pf}\left(
    \frac{(z_i-\xi_1)(z_j-\xi_2)(z_i-\xi_3)(z_j-\xi_4)
      +(z_i\leftrightarrow z_j)}{z_i-z_j}
  \right).
\end{align}
We see that $\xi_1$ and $\xi_3$ belong to one group in that they constitute additional zeros seen by one member of each electron pair, while $\xi_2$ and $\xi_4$ belong to another group as they constitute zeros seen by the other members of each electron pair.  The wave function is symmetric (or antisymmetric, depending on the number of electron pairs) under interchange of both groups.  Since the internal Hilbert space is spanned by the distinct possibilities of affiliating the individual quasiholes with the two groups, the state in this space will change as we adiabatically interchange two quasiholes belonging to different groups, say $\xi_3$ and $\xi_4$ in \eqref{eq:pf4QH}.  Naively, one might think that the dimension of the internal Hilbert space is given by the number of ways to partition the quasiholes at $\xi_1,\ldots,\xi_{2n}$ into two different groups, \ie by $(2n-1)!!$ for $2n$ quasiholes.  Note that the number of quasiholes has to be even on closed surfaces to satisfy the Dirac flux quantization condition.  The true dimension of the internal Hilbert space, however, is only $2^{n-1}$\cite{nayak-96npb529}.  
The reason for this is that the internal Hilbert space is spanned by Majorana fermion states in the vortex cores\cite{read-00prb10267}, as we will elaborate in the following subsection.

The statistics is non-Abelian in the sense that the order according to
which we interchange quasiholes matters.  Let the unitary matrix $M_{ij}$
describe the rotation of the internal Hilbert space state vector which
describes the adiabatic interchange of two quasiholes at $\xi_i$ and
$\xi_j$:
\begin{equation*}
  \label{eq:pfMij}
  \ket{\psi}\to M_{ij}\ket{\psi}.
\end{equation*}
The statistics is non-Abelian if the matrices associated with
successive interchanges do not commute in general,
\begin{equation*}
  M_{ij}M_{jk}\ne M_{jk} M_{ij}.
\end{equation*}
Note that the internal state vector is protected in the sense that it
is insensitive to local perturbations---it can \emph{only} be
manipulated through braiding of the vortices.  For a sufficiently
large number of vortices, on the other hand, any unitary
transformation in this space can be approximated to arbitrary accuracy
through successive braiding operations\cite{freedman-02cmp587}.
These properties together render non-Abelions preeminently suited for
applications as protected qubits in quantum
computation \cite{nayak-08rmp1083,stern10n187}.

\subsection*{Majorana fermions in vortex cores}

The key to understanding the non-Abelian statistics 
of the quasiparticle excitations of the Pfaffian state lies in the
Majorana fermion modes in the vortices of p-wave
superfluids
\cite{read-00prb10267}. 
The p-wave pairing symmetry implies that the order parameter for the
superfluid acquires a phase of $2\pi$ as we go around the Fermi
surface,
\begin{align}
  \label{eq:pfpwaveOP}
  \langle {c_{\bs{k}}^\dagger\,c_{-\bs{k}}^\dagger} \rangle
  =\Delta_0(k)\cdot(k_\x+\text{i}k_\y),
\end{align}
where $\Delta_0(k)$ can be chosen real.  
The Hamiltonian for a single vortex at the origin is given by
\begin{align}
  \label{eq:pfHvortex}
  H=\int\!\text{d}\bs{r}
  \left\{\psi^{\dagger}\!\left(\!-\frac{\bs{\nabla}^2}{2m}-
       \varepsilon_F\!
    \right)\!\psi
    +\psi^{\dagger}\left(
      e^{\text{i}\varphi}\Delta_0(r)*(\partial_x-\text{i}\partial_y)
    \right)\psi^{\dagger}+\text{h.c.}\right\},
\end{align}
where $A*B\equiv\frac{1}{2}\{A,B\}$ denotes the symmetrized product, and $r$ and $\varphi$ are polar coordinates.  The order parameter $\Delta_0(r)$ vanishes inside the vortex core.  We can obtain the energy eigenstates localized inside the vortex by solving the Bogoliubov--de Gennes equations\cite{deGennes66}
\begin{align}
  \label{eq:pfBogodeGennes}
  \comm{H}{\gamma_n^\dagger(\bs{x})}=E_n\gamma_n^\dagger(\bs{x}),
\end{align}
where $n$ labels the modes and 
\begin{align}
  \label{eq:pfBogoOp}
  \gamma_n^\dagger(\bs{x})
  =u_n(\bs{x})\psi^\dagger(\bs{x})+v_n(\bs{x})\psi(\bs{x})
\end{align}
are the Bogoliubov quasiparticle operators.  The low energy spectrum
is given by 
\begin{align}
  \label{eq:pfEn}
  E_n=n\omega_0,
\end{align}
where $n$ is an integer and $\omega_0=\Delta^2/\varepsilon_F$ the
level spacing.  Note that while in an s-wave superfluid, the
Bogoliubov operators 
\begin{align}
  \label{eq:pfBogoOpswave}
  \gamma_{n\up}^\dagger(\bs{x})
  =u_{n\up}(\bs{x})\psi^\dagger(\bs{x})+v_{n\dw}(\bs{x})\psi(\bs{x})
\end{align}
combine $\up$-spin electron creation operators with $\dw$-spin annihilation operators, in the p-wave superfluid, the operators \eqref{eq:pfBogoOp} combine creation and annihilation
operators of the same spinless (or spin-polarized) fermions.  Since the Bogoliubov--de Gennes equations are not able to distinguish between particles and antiparticles, we obtain each physical solution twice:  once with positive energy as a solution of the
Bogoliubov--de Gennes equation \eqref{eq:pfBogodeGennes} for the creation operators, and once with negative energy as a solution of the same equation for the annihilation operators,
\begin{align}
  \label{eq:pfBogodeGennesann}
  \comm{H}{\gamma_n(\bs{x})}=-E_n\gamma_n(\bs{x}),
\end{align}
which is obtained from \eqref{eq:pfBogodeGennes} by Hermitian conjugation.  We resolve this technical artifact by discarding the negative energy solutions as unphysical.  For the $n=0$ solution at $E_0=0$, it implies that we get one fermion solution when we overcount by a factor of two.  The physical solution at $E=0$ is hence given by one half of a fermion, or a Majorana fermion, as
\begin{align}
  \label{eq:pfMajo1}
  \gamma_0^\dagger(\bs{x})=\gamma_0(\bs{x}).
\end{align}

In general, one fermion $\psi,\psi^\dagger$ consists of two Majorana 
fermions,
\begin{align}
  \label{eq:pfPsiGamma}
  \psi=\frac{1}{2}(\gamma_1+\text{i}\gamma_2),\qquad
  \psi^\dagger=\frac{1}{2}(\gamma_1-\text{i}\gamma_2),
\end{align}
which in turn are given by the real and imaginary part of the fermion operators,
\begin{align}
  \label{eq:pfGammaPsi}
  \gamma_1
  =\psi +\psi^\dagger,\qquad
  \gamma_2
  =-\i(\psi -\psi^\dagger).
\end{align}
They obey the anticommutation relations
\begin{align}
  \label{eq:pfMajoComm}
  \anticomm{\gamma_i}{\gamma_j}=2\delta_{ij},
\end{align}
as one may easily verify with \eqref{eq:pfGammaPsi}.  Majorana fermions are their own antiparticles, as $\gamma_i^\dagger=\gamma_i$.  If we write the basis for a single fermion as $\{\vac,\psi^\dagger\vac\}$, we can write the fermion creation and
annihilation operators as 
\begin{align}
  \label{eq:pf2compbasis}
  \psi^\dagger=\!\left(\!\!
    \begin{array}{cc}
      \,0\,&\,0\,\\1&0 
    \end{array}\!\!\right),\quad
  \psi=\!\left(\!\!
    \begin{array}{cc}
      \,0\,&\,1\,\\0&0 
    \end{array}\!\!\right).
\end{align}
In this basis, the Majorana fermions are given by the first two Pauli
matrices,
\begin{align}
  \label{eq:pf2compbasis2}
  \gamma_1=\!\left(\!\!
    \begin{array}{cc}
      \,0\,&\,1\,\\1&0 
    \end{array}\!\!\right)=\sigma_\x,\quad
  \gamma_2=\!\left(\!\!
    \begin{array}{cc}
      \,0\,&\,-\i\,\\\i&0 
    \end{array}\!\!\right)=\sigma_\y.
\end{align}

\begin{figure}[bt] 
  \begin{center}
  \begin{tikzpicture}[>=latex,scale=1]
  \begin{scope}[shift={(0,0)}]
    \draw[thick](0,-2)--(0,3) node[below right] {\small boundary};
    \draw[thick,dotted]
    (0,1.5)--node[above]{\small branch cut}(2.94,1.5);       
    \draw[thick]( 3,1.5) 
    circle[radius=.06] node[right]{\small $\gamma_{i+1}$};
    \draw[thick,dotted](0,0)--(2.64,0);       
    \draw[thick,fill]( 2.7,0) 
    circle[radius=.05] node[right]{\small $\gamma_{i}$};
    \begin{scope}[shift={(3,1.5)},rotate=100]
      \draw[thin,->](0,.06) arc[start angle=2,end angle=117.5,radius=2];
      \draw[thick,fill](-3,{sqrt(3)}) circle[radius=.05];
    \end{scope}
    \draw[thin,<-] 
    (1.5,-.1)--(2.4,-1)node[right]
      {\small as $\gamma_{i+1}$ crosses the};
    \node[right] at (2.4,-1.5)
      {\small branch cut: $\gamma_{i+1}\to-\gamma_{i+1}$};
  \end{scope}
  \begin{scope}[shift={(8,0)},xscale=1,yscale=1]
    \draw[thin,->](0,-1.6)--(0,1.6) node[above] {\small time};
    \begin{scope}[shift={(1.7,{-sqrt(3)/2})}]
      \braiding
      \node[right] at (0,{sqrt(3)/2}){\small $T_{i}$};  
      \node[below] at(-1,0){\small $\gamma_{i}$};
      \node[below] at (0,0){\small $\gamma_{i+1}$};    
    \end{scope}
  \end{scope}
  \end{tikzpicture}
  \end{center}
  \caption{The Majorana fermion $\gamma_{i+1}$ acquires a $-$ sign as it crosses the branch cut from another vortex.} 
  \label{fig:pfBranchCut}
\end{figure}
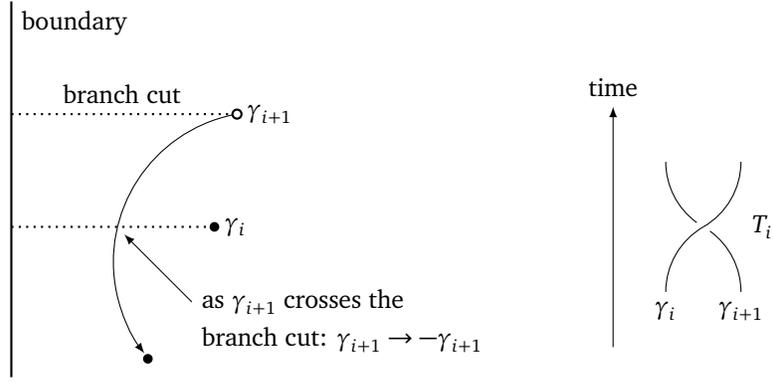

Returning to vortices in a p-wave superfluid, note that by definition, the order parameter acquires a phase of $2\pi$ as we go around a vortex.  This implies that the electron creation and annihilation operators acquire a phase $\pi$, or a minus sign, which implies via \eqref{eq:pfGammaPsi} that the Majorana fermion states also acquire a minus sign,
\begin{align}
  \label{eq:pfMajoMinus}
  \gamma_i\to -\gamma_i,
\end{align}
as we encircle a vortex.  By choice of gauge, we can implement the phase change of $2\pi$ in the superconducting order parameter as a branch cut connecting the vortices to the left boundary of the system, and assume a convention according to which the Majorana fermion in each vortex crossing a branch cut acquires a minus sign, as illustrated in Figure \ref{fig:pfBranchCut}.

\subsection*{Statistics of Ising anyons}

To obtain the non-Abelian statistics, Ivanov\cite{ivanov01prl268} constructed a representation $\tau(T_i)$ of the braid group $B_{2n}$ on $\mathbb{R}^2$ with the algebra \eqref{eq:pfBraidAlgebra}, such that, according to the convention defined in Figure \ref{fig:pfBranchCut}, only Majorana fermion $\gamma_{i+1}$ acquires a minus sign under the braiding operation $T_i$. Then the transformation rule is  \begin{align}
  \label{eq:pfTransRule}
  T_i(\gamma_j)=\left\{\begin{array}{@{}ll}
                         \phantom{-}\gamma_{j+1} &\text{for}\ i=j,\\
                         -\gamma_{j-1} &\text{for}\ i=j-1,\\
                         \phantom{-}\gamma_{j} &\text{otherwise}.
                       \end{array}\right.
\end{align}
To describe the action of these transformations on the (internal) state vectors, we hence need to find a representation $\tau(T_i)$ of the braid group $B_{2n}$ such that 
\begin{align}
  \label{eq:pfRep}
  \tau(T_i)\,\gamma_{j}\,\tau(T_i)^{-1}= T_i(\gamma_j)
\end{align}
with $T_i(\gamma_j)$ given by \eqref{eq:pfTransRule}.  The
solution is\cite{ivanov01prl268} 
\begin{align}
  \label{eq:pfRepSolution}
  \tau(T_i)&=\exp\left(\frac{\pi}{4}\gamma_{i+1}\gamma_{i}\right)
  \nonumber\\[0.2\baselineskip]
  &=\cos\left(\frac{\pi}{4}\right) 
  + \gamma_{i+1}\gamma_{i}\,\sin\left(\frac{\pi}{4}\right),
  \nonumber\\
  &=\frac{1}{\sqrt{2}}(1+\gamma_{i+1}\gamma_{i}),
\end{align}
as one can easily verify using $(\gamma_{i+1}\gamma_{i})^2=-1$.  The
inverse transformation is given by
\begin{align}
  \label{eq:pfRepSolution-1}
  \tau(T_i)^{-1} 
  =\frac{1}{\sqrt{2}}(1-\gamma_{i+1}\gamma_{i}).
\end{align}
A few steps of algebra yield
\begin{equation*}
  \tau(T_1)
  \left\{\begin{array}{@{}c@{}}\gamma_1\\\gamma_2 \end{array}\right\}
  \tau(T_1)^{-1}
  =\left\{\begin{array}{@{}r@{}}\gamma_2\\-\gamma_1 \end{array}\right\}.
\end{equation*}

The simplest examples of this representation are the cases of two and four vortices\cite{nayak-96npb529,ivanov01prl268}, 
which we will elaborate now.  In the case of two vortices, the two Majorana fermions $\gamma_1$ and $\gamma_2$ can be combined into a single fermion via \eqref{eq:pfPsiGamma}, and the ground state is two-fold degenerate.  The braid group $B_2$ has only one generator $T_1$ with representation 
\begin{align}
  \label{eq:pfRepT1}
  \tau(T_1)&=\exp\left(\frac{\pi}{4}\gamma_{2}\gamma_{1}\right)
  \nonumber\\[0.2\baselineskip]
  &=\exp\left(-\text{i}\frac{\pi}{4}
    (\psi-\psi^\dagger)(\psi+\psi^\dagger)\right)
  \nonumber\\[0.2\baselineskip]
  &=\exp\left(\text{i}\frac{\pi}{4}
    (2\psi^\dagger\psi-1)\right)
  \nonumber\\[0.2\baselineskip]
  &=\exp\left(-\text{i}\frac{\pi}{4}\sigma_\z \right),
\end{align}
where $\sigma_\z $
is the third Pauli matrix in the basis
$\big\{\vac,\psi^\dagger\vac\big\}$.  The braiding is hence diagonal in this
basis, and only gives an overall phase, which depends on whether the
fermion state is occupied or not.

The non-Abelian statistics manifests itself only once we consider four
vortices.  Following Ivanov\cite{ivanov01prl268}, we combine the
four Majorana fermions into two fermions,
\begin{align}
  \label{eq:pfPsiGamma1234}
  \psi_1=\frac{1}{2}(\gamma_1+\text{i}\gamma_2),\qquad
  \psi_2=\frac{1}{2}(\gamma_3+\text{i}\gamma_4),
\end{align}
with similar expressions for the fermion creation operators $\psi_1^\dagger$,
$\psi_2^\dagger$.  The braid group $B_4$ has three generators $T_1$,
$T_2$, and $T_3$.  Their representations in a basis of fermion
occupation numbers
\begin{equation*}
  \big\{\vac,
  \psi_1^\dagger\vac,\psi_2^\dagger\vac,\psi_1^\dagger\psi_2^\dagger\vac\big\}
\end{equation*}
are given by two diagonal operators
\begin{align}
  \tau(T_1)&= 
  \exp\left(\frac{\pi}{4}\gamma_{2}\gamma_{1}\right)=
  \exp\left(-\text{i}\frac{\pi}{4}{\sigma_\z^{(1)}}\right)
  =\left(\!\!\begin{array}{cccc}
       e^{-\i\pi/4}&0&0&0\\ 
       0& e^{\i\pi/4}&0&0\\ 
       0&0&e^{-\i\pi/4}&0\\
       0&0&0& e^{\i\pi/4}
     \end{array}\!\!\right),
  \nonumber
\end{align}
\begin{align}
  \tau(T_3)&= 
  \exp\left(\frac{\pi}{4}\gamma_{4}\gamma_{3}\right)=
  \exp\left(-\text{i}\frac{\pi}{4}{\sigma_\z^{(2)}}\right)
  =\left(\!\!\begin{array}{cccc}
       e^{-\i\pi/4}&0&0&0\\ 
       0&e^{-\i\pi/4}&0&0\\ 
       0&0& e^{\i\pi/4}&0\\
       0&0&0& e^{\i\pi/4}
     \end{array}\!\!\right),
  \nonumber
\end{align}
and one off-diagonal operator
\begin{align}
  \tau(T_2)&= 
  \exp\left(\frac{\pi}{4}\gamma_{3}\gamma_{2}\right)
  \nonumber\\ 
  &= 
  \frac{1}{\sqrt{2}}
  \left(1-\i(\psi_2+\psi_2^\dagger)(\psi_1-\psi_1^\dagger)\right)
  =
  \left(\!\!\begin{array}{cccc}
       1&0&0&\,-\i\,\\ 
       0&1&\,-\i\,&0\\ 
       0&\,-\i\,&1&0\\ 
       \,-\i\,&0&0&1
     \end{array}\!\!\right).
  \nonumber
\end{align}

Note that since the representations $\tau(T_i)$ given by \eqref{eq:pfRepSolution} are even in fermion parity (\ie they change the fermion numbers only by even integers), we may restrict them to only even or odd sectors in the fermion numbers.  For the example of four vortices, these sectors are given by $\{\vac,\psi_1^\dagger\psi_2^\dagger\vac\}$ and $\{\psi_1^\dagger\vac,\psi_2^\dagger\vac\}$.  Each sector contains $2^{n-1}$ states, which is the degeneracy found for a Pfaffian state with an even number of electrons\cite{nayak-96npb529}.  Physically, this reflects that while the number of fermions is not a good quantum number in a superfluid, the fermion parity is a good quantum number.
  
Finally, note that the derivation of the non-Abelian statistics depends only on (a) the 
existence of Majorana fermion modes in the vortex cores, and (b) the Majorana fermions changing sign $\gamma_i\to -\gamma_i$ when the phase of the superfluid order parameter changes by $2\pi$, as it does by definition when we go around a vortex.


The Ising anyons elaborated here constitute the simplest kind of non-Abelian anyons.  They are supported by the Pfaffian state, which is also the simplest (\ie $k=2$) non-Abelian quantum Hall state of the Read--Rezayi series\cite{Read-99prb8084},
\begin{align}
  \label{eq:psirr}
  \psi^{\text{RR}}_k(z_1,\dots,z_{N})
  =\mathcal{S}\left\{\prod_{m=1}^{k}\left(
    \prod_{\substack{i,j=(m-1)M+1\\[1pt] i<j}}^{mM}(z_i-z_j)^2 
  \right) \right\}\, 
  \prod_{i<j}^{N} (z_i-z_j),
\end{align}
where $\mathcal{S}$ denotes symmetrization, and the number of electrons $N=mM$ has to be a multiple of the integer $M$.  The filling fraction of \eqref{eq:psirr} is given by $\nu=\frac{k}{k+2}$.  The equivalence of the $k=2$ state with the Pfaffian state \eqref{eq:pfpsi0} can be shown easily with an identity due to Frobenius\cite{greiter-92npb567}.
 
For topological quantum computing, the prime candidates are Fibonacci anyons\cite{trebst-08ptps384}, as these are the simplest candidates suited for universal quantum computation, as they are capable of simulating any program any other quantum computer can carry out\cite{nayak-08rmp1083,bonesteel23proc}.  These can be realized by the $k=3$ state of the Read--Rezayi series, which may be realized as the particle-hole conjugate of the observed plateau at $\nu=12/5$ in the second Landau level, in spin-singlet generalizations \cite{ardonne-99prl5096} of those, or in certain classes of bilayer quantum Hall states\cite{vaezi-14prl236804}.

The spinon excitations of spin chains with spin $S=1$ and higher tuned to multi-critical points provide examples of non-Abelian anyons in 1D\cite{bouwknegt-99npb501,greiter-19prb115107}.  One way to understand this is that the higher spin $S$ chains can be described by Read--Rezayi states with $k=2S$ for renormalized spin flip operators acting on the vacuum where the spins are maximally polarized, which renders the spinon excitations both conceptually and formally similar to the non-Abelian, fractionally charged quasiparticles of the Read--Rezayi states.  The fusion rules for the anyons are then likewise given by the graded algebra su(2) level $k$, and the internal space of the non-Abelian anyons is spanned by configurations of topological shifts between the single-particle momenta of the spinons.  So while the spinons in the spin $\half$ HSM are similar to the quasiparticle excitations of Laughlin states, the spinons in multi-critical states in higher spin chains are similar to the non-Abelian quasiparticle excitations of Read--Rezayi states.

\section{Experimental observations of anyons}

The experimental situation around fractional statistics and anyons is evolving rapidly.  As we will indicate in our concluding section, there are many directions being actively explored.   In this section we will focus on a few fundamental experiments that exhibit the existence of unconventional quantum statistics in an especially clear and convincing way.   Those experiments are broadly of two kinds: experiments in fractional quantum Hall states, and experiments in engineered systems.

The unconventional quantum statistics of quasiparticles in fractional quantum Hall states is a rigorous consequence of a theory whose principles have been tested experimentally in many ways, as we have seen.  In this sense there has long been overwhelming indirect evidence for the existence of the predicted anyons: It is hard---though maybe not impossible---to articulate an alternative theory that is even qualitatively consistent with experiments and yet does not contain anyons.  Nevertheless, it has been a major goal of experimental work to exhibit the most characteristic anyon behaviors, such as phased interference between alternate paths and braiding, directly.  This is of course fundamentally interesting in itself, and the techniques being developed to achieve it promise to have wide application. 

Two recent breakthrough experiments on the $\nu = 1/3$ quantum Hall state have achieved this goal.  Since these experiments have dedicated articles within this {\it Encyclopedia}, our discussion here will be brief and broadly conceptual. 


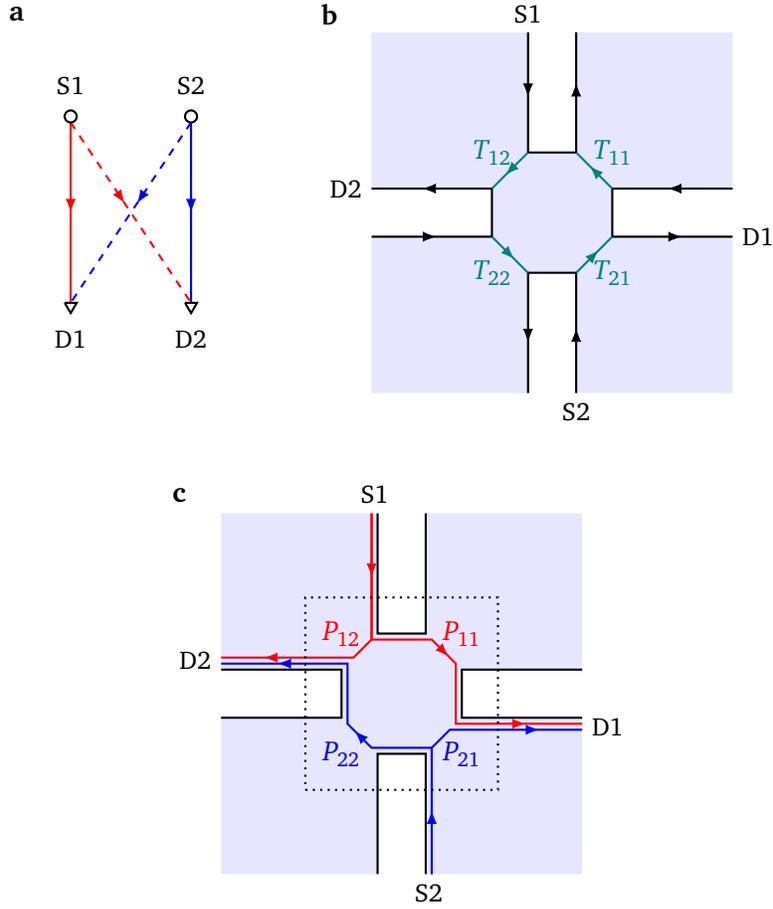
\begin{figure}[t]
\begin{center}
\begin{tikzpicture}[>=latex,scale=.80]
\coordinate (Origin)   at (0,0);
\pgfmathsetmacro{\xd}{1}
\pgfmathsetmacro{\yd}{1.5}
\pgfmathsetmacro{\cd}{.1}
\pgfmathsetmacro{\rd}{.1}
%
%
\begin{scope}[shift={(-4,8)}]
  \def\centercolor{teal}
  \begin{scope}[decoration={markings,mark=at position .5 with {\arrow{>}}}] 
    \draw[red,thick,postaction={decorate}](-\xd,\yd)--(-\xd,-\yd);
    \draw[blue,thick,postaction={decorate}](\xd,\yd)--(\xd,-\yd);
  \end{scope}
  \begin{scope}[decoration={markings,mark=at position .45 with {\arrow{>}}}] 
    \draw[red,dashed,thick,postaction={decorate}](-\xd,\yd)--(\xd,-\yd);
    \draw[blue,dashed,thick,postaction={decorate}](\xd,\yd)--(-\xd,-\yd);
  \end{scope}
  \draw [thick] (-\xd,\yd+\rd) circle [radius=\rd];  
  \draw [thick] (\xd,\yd+\rd) circle [radius=\rd];  
  \draw [thick] (-\xd-\rd,-\yd)--(-\xd,-\yd-1.71*\rd)--(-\xd+\rd,-\yd)--(-\xd-\rd,-\yd);
  \draw [thick] (\xd-\rd,-\yd)--(\xd,-\yd-1.71*\rd)--(\xd+\rd,-\yd)--(\xd-\rd,-\yd);
  \node [above] at (-\xd,\yd+3*\rd) {\small S1}; 
  \node [above] at (\xd,\yd+3*\rd) {\small S2}; 
  \node [below] at (-\xd,-\yd-3*\rd) {\small D1}; 
  \node [below] at (\xd,-\yd-3*\rd) {\small D2}; 
  \node at (-1.9,3.3) {\bf a}; 
\end{scope}
%
%
\begin{scope}[shift={(3,8)}]
  \def\centercolor{teal}
  \path[fill=blue!10!white] 
  (3,3)--(.4,3)--(.4,1)--(1,.4)--(3,.4)--(3,3);
  \path[fill=blue!10!white,rotate=90] 
  (3,3)--(.4,3)--(.4,1)--(1,.4)--(3,.4)--(3,3);
  \path[fill=blue!10!white,rotate=180] 
  (3,3)--(.4,3)--(.4,1)--(1,.4)--(3,.4)--(3,3);
  \path[fill=blue!10!white,rotate=270] 
  (3,3)--(.4,3)--(.4,1)--(1,.4)--(3,.4)--(3,3);
  \path[fill=blue!10!white] (1,.4)--(.4,1)--(-.4,1)--(-1,.4)
  --(-1,-.4)--(-.4,-1)--(.4,-1)--(1,-.4)--(1,.4);
  \begin{scope}[decoration={markings,
      mark=at position .225 with {\arrow{>}},
      mark=at position .825 with {\arrow{>}}}] 
    \draw[thick,postaction={decorate}]
    (-.4,3)--(-.4,1)--(.4,1)--(.4,3);
    \draw[thick,postaction={decorate},rotate=90]
    (-.4,3)--(-.4,1)--(.4,1)--(.4,3);
    \draw[thick,postaction={decorate},rotate=180]
    (-.4,3)--(-.4,1)--(.4,1)--(.4,3);
    \draw[thick,postaction={decorate},rotate=270]
    (-.4,3)--(-.4,1)--(.4,1)--(.4,3);
  \end{scope}
  \begin{scope}[decoration={markings,
      mark=at position .625 with {\arrow{>}}}] 
    \draw[\centercolor,thick,postaction={decorate}]
    (-.4,1)--(-1,.4);
    \draw[\centercolor,thick,postaction={decorate},rotate=90]
    (-.4,1)--(-1,.4);
    \draw[\centercolor,thick,postaction={decorate},rotate=180]
    (-.4,1)--(-1,.4);
    \draw[\centercolor,thick,postaction={decorate},rotate=270]
    (-.4,1)--(-1,.4);
  \end{scope}
  \node [above] at (-.4,3) {\small S1}; 
  \node [left]  at (-3,.4) {\small D2}; 
  \node [below] at (.4,-3) {\small S2}; 
  \node [right] at (3,-.4) {\small D1}; 
  \node [\centercolor] at (1,1)  {$T_{11}$};
  \node [\centercolor] at (-1,1)  {$T_{12}$};
  \node [\centercolor] at (-1,-1)  {$T_{22}$};
  \node [\centercolor] at (1,-1)  {$T_{21}$};
  \node at (-3.7,3.3) {\bf b}; 
\end{scope}
%
%
\begin{scope}[shift={(0.5,0)}]
  \def\centercolor{teal}
  \path[fill=blue!10!white] 
  (3,3)--(.4,3)--(.4,1)--(1,.4)--(3,.4)--(3,3);
  \path[fill=blue!10!white,rotate=90] 
  (3,3)--(.4,3)--(.4,1)--(1,.4)--(3,.4)--(3,3);
  \path[fill=blue!10!white,rotate=180] 
  (3,3)--(.4,3)--(.4,1)--(1,.4)--(3,.4)--(3,3);
  \path[fill=blue!10!white,rotate=270] 
  (3,3)--(.4,3)--(.4,1)--(1,.4)--(3,.4)--(3,3);
  \path[fill=blue!10!white] (1,.4)--(.4,1)--(-.4,1)--(-1,.4)
  --(-1,-.4)--(-.4,-1)--(.4,-1)--(1,-.4)--(1,.4);
  \begin{scope}
    \draw[thick](-.4,3)--(-.4,1)--(.4,1)--(.4,3);
    \draw[thick,rotate=90](-.4,3)--(-.4,1)--(.4,1)--(.4,3);
    \draw[thick,rotate=180](-.4,3)--(-.4,1)--(.4,1)--(.4,3);
    \draw[thick,rotate=270](-.4,3)--(-.4,1)--(.4,1)--(.4,3);
  \end{scope}
  \begin{scope}[decoration={markings,
      mark=at position .225 with {\arrow{>}},
      mark=at position .85 with {\arrow{>}}}] 
    \def\currentcolor{red}
    \draw[\currentcolor,thick,postaction={decorate}]
    (-.4-\cd,3)--(-.4-\cd,1-\cd)--(-1+2*\cd,.4+2*\cd)--(-3,.4+2*\cd);
    \node[\currentcolor] at (-1,1)  {$P_{12}$};
    \def\currentcolor{blue}
    \draw[\currentcolor,thick,postaction={decorate},rotate=180]
    (-.4-\cd,3) --(-.4-\cd,1-\cd)--(-1+2*\cd,.4+2*\cd)--(-3,.4+2*\cd);
    \node[\currentcolor] at (1,-1)  {$P_{21}$};
  \end{scope}
  \begin{scope}[decoration={markings,
      mark=at position .3 with {\arrow{>}}, 
      mark=at position .8 with {\arrow{>}}}] 
    \def\currentcolor{red}
    \draw[\currentcolor,thick,postaction={decorate}]
    (-.4-\cd,1-\cd)--(.4+\cd,1-\cd)--(1-\cd,.4+\cd)--(1-\cd,-.4-\cd)--(3,-.4-\cd);
    \draw[\currentcolor,thick] (-.4-\cd,3)--(-.4-\cd,1-\cd);
    \node[\currentcolor] at (1,1)  {$P_{11}$};
    \def\currentcolor{blue}
    \draw[\currentcolor,thick,postaction={decorate},rotate=180]
    (-.4-\cd,1-\cd)--(.4+\cd,1-\cd)--(1-\cd,.4+\cd)
    --(1-\cd,-.4-\cd)--(3,-.4-\cd);
    \draw[\currentcolor,thick ,rotate=180] (-.4-\cd,3)--(-.4-\cd,1-\cd);
    \node[\currentcolor] at (-1,-1)  {$P_{22}$};
  \end{scope}
  \draw [dotted,thick] (-1.6,-1.6) rectangle (1.6,1.6);
  \node [above] at (-.4-.5*\cd,3) {\small S1}; 
  \node [left]  at (-3,.4+1.5*\cd) {\small D2}; 
  \node [below] at (.4+.5*\cd,-3) {\small S2}; 
  \node [right] at (3,-.4-1.5*\cd) {\small D1}; 
  \node at (-3.7,3.3) {\bf c}; 
\end{scope}
\end{tikzpicture}
\end{center}
\caption{Schematic depiction of the ``anyon collider'' experiment, emphasizing its kinship with intensity interferometry.   {\bf a}. Principle of the classic intensity interferometer.  {\bf b}.  Edge mode propagation from sources to detectors through weak links.  {\bf c}.  Alternative paths for production of correlated excitations.  
} 
\label{fig:feve}
\end{figure}

The experiment of Bartolomei \etal\cite{bartolomei-20s173} is has been described as an ``anyon collider''.  It can also be considered to be a form of intensity interferometer, such as we meet in the Hanbury Brown--Twiss (HBT) effect\cite{hanburybrown-56n1046}.   Its central idea is depicted schematically in Figure~\ref{fig:feve}.  Figure~\ref{fig:feve}a recalls the paradigmatic HBT setup as it was originally developed, in astronomy.  One has two sources S1, S2 of photons, and two detectors D1, D2, and measures correlations between the signals received in the detectors as a function of time, and of their separation.  Non-trivial correlations in the noise amplitudes detected simultaneously at D1, D2 can arise because of interference between two distinct ways they can be triggered.  Indeed, as indicated in Figure~\ref{fig:feve}a, there is a ``direct'' contribution from the two-photon process indicated by the solid lines, and an ``exchange'' contribution arising from the two-photon process indicated by the dotted lines.  The rule for combining those process---namely, to add the products of propagators, and then square to obtain the probability---reflects both the free propagation factors and the Bose statistics of photons.  This philosophy has been widely generalized and exploited: in quantum optics, where one guides the photons using mirrors and beam-splitters, in heavy ion collisions, where where pairs of pions or other particles are detected, and previously in condensed matter physics with guided electrons\cite{baym98appb1839}.  

Figure~\ref{fig:feve}b indicates some of the special features that arise in the anyon collider.  The light blue background represents a bulk $\nu = 1/3$ fractional Hall state. 
It has four metallic contacts, are indicated by the while areas. 
One has sources S1 and S2 of anyons, and electric current detectors D1, D2.  (The anyons are produced by injecting electrons at additional contacts, not indicated here, and collecting current that passes through weak links, necessarily in the form of anyon quasiparticles.)   Current is carried by chiral modes that propagate along the boundary of the Hall fluid.  They are unidirectional, such that if a right thumb is laid down following the direction, the index finger will point into the fluid.   One sees that propagation from S1 to D2 involves tunneling at the weak link $T_{12}$, and similarly for the other ways of connecting sources to detectors.  

Figure~\ref{fig:feve}c brings the ideas of \ref{fig:feve}a and \ref{fig:feve}b together.  One sees that there are two red paths that take anyons from S1 to D1 and D2, analogous to the photon paths we contemplated before; that the contacts act to guide their propagation; and that the weak links act as beam splitters.  Also indicated here is the possibility of defining a ``reaction region'', indicated by the dotted line.  If we treat this region as a black box, then we can visualize  the whole set-up as a collider surrounded by detectors. 

The actual calculation of propagation factors for quantum Hall quasiparticles is technically challenging, and involves modeling of the edge states.  The model used for the analysis\cite{rosenow-16prl156801}, which was proposed before the experiment, works well to describe a wealth of measurements.  Using it, one can extract the statistics of the quasiparticles, which appears as an independent parameter.  The predicted value of the fractional statistics parameter, $\phi = \pi/3$, is consistent with the experimentally observed data, whose precision is $\lesssim 10 \%$.  Boson or fermion statistics is very clearly excluded.  While this is quite convincing, it would be good to find a protocol based on HBT that in principle converges on the exact statistical phase, as we do have for the fractional charge and the Hall coefficient.  


\begin{figure}[tb]
\begin{center}
\begin{tikzpicture}[>=latex,scale=.8]
\coordinate (Origin)   at (0,0);
\pgfmathsetmacro{\xd}{1}
\pgfmathsetmacro{\yd}{1.5}
\pgfmathsetmacro{\cd}{.1}
\pgfmathsetmacro{\rd}{.1}
\begin{scope}[shift={(0,0)}]
  \def\centercolor{teal}
  \path[fill=blue!10!white] (.3,3)--(.3,1.4)--(1.4,1.4)--(1.4,.8)--(0,.8)
     --(0,0)--(2.8,0)--(2.8,3)--(2.1,3)--(2.1,.4)--(1.5,.4)--(1.5,3);
  \path[fill=blue!10!white,rotate=180]
  (.3,3)--(.3,1.4)--(1.4,1.4)--(1.4,.8)--(0,.8)
     --(0,0)--(2.8,0)--(2.8,3)--(2.1,3)--(2.1,.4)--(1.5,.4)--(1.5,3);
  \begin{scope}[xscale=-1,yscale=1]
  \path[fill=blue!10!white] (.3,3)--(.3,1.4)--(1.4,1.4)--(1.4,.8)--(0,.8)
     --(0,0)--(2.8,0)--(2.8,3)--(2.1,3)--(2.1,.4)--(1.5,.4)--(1.5,3);
  \path[fill=blue!10!white,rotate=180]
  (.3,3)--(.3,1.4)--(1.4,1.4)--(1.4,.8)--(0,.8)
     --(0,0)--(2.8,0)--(2.8,3)--(2.1,3)--(2.1,.4)--(1.5,.4)--(1.5,3);
  \end{scope}
  \begin{scope}
    \draw[thick](-.3,3)--(-.3,1.4)--(-1.4,1.4)--(-1.4,.8)
              --(1.4,.8)--(1.4,1.4)--(.3,1.4)--(.3,3);
    \draw[thick,rotate=180](-.3,3)--(-.3,1.4)--(-1.4,1.4)--(-1.4,.8)
              --(1.4,.8)--(1.4,1.4)--(.3,1.4)--(.3,3);
  \end{scope}
  \begin{scope}[shift={(1.8,0)},decoration={markings,
      mark=at position .225 with {\arrow{>}},
      mark=at position .825 with {\arrow{>}}}] 
    \draw[thick,postaction={decorate}](-.3,3)--(-.3,.4)--(.3,.4)--(.3,3);
    \draw[thick,postaction={decorate},rotate=180]
      (-.3,3)--(-.3,.4)--(.3,.4)--(.3,3);
    \draw[\centercolor,thick](0,.4)--(0,-.4);
  \end{scope}
  \begin{scope}[shift={(-1.8,0)},decoration={markings,
      mark=at position .225 with {\arrow{>}},
      mark=at position .825 with {\arrow{>}}}] 
    \draw[thick,postaction={decorate}](-.3,3)--(-.3,.4)--(.3,.4)--(.3,3);
    \draw[thick,postaction={decorate},rotate=180]
      (-.3,3)--(-.3,.4)--(.3,.4)--(.3,3);
    \draw[\centercolor,thick](0,.4)--(0,-.4);
  \end{scope}
  \begin{scope}[decoration={markings,
      mark=at position .365 with {\arrow{>}},
      mark=at position .648 with {\arrow{>}}}] 
    \def\currentcolor{red}
    \draw[\currentcolor,thick,postaction={decorate}]
    (2.1+\cd,-3)--(2.1+\cd,-.4+\cd)--(1.5-\cd,-.4+\cd)--(1.5-\cd,-.8+\cd)
    --(-1.5+\cd,-.8+\cd)--(-1.5+\cd,.8-\cd)--(1.5-\cd,.8-\cd)
    --(1.5-\cd,.4-\cd)--(2.1+\cd,.4-\cd)--(2.1+\cd,3);
  \end{scope}
  \begin{scope}[decoration={markings,
      mark=at position .52 with {\arrow{>}}}] 
    \def\currentcolor{red}
    \draw[\currentcolor,thick,postaction={decorate}]
    (2.1+2*\cd,-3)--(2.1+2*\cd,3);
  \end{scope}
  \node [above] at (2.1+1.5*\cd,3) {\small D};
  \node [below] at (2.1+1.5*\cd,-3) {\small S}; 
\end{scope}
\end{tikzpicture}
\end{center}
\caption{Schematic depiction of the ``braiding'' experiment, emphasizing its kinship with Fabry-Perot interferometry.  Weak links supply beam-splitters for anyons propagating as edge modes.  The two alternative paths from source to detector differ in surrounding, or not, the central island.  Their interference is sensitive to the presence of bulk anyons within the island.  
} 
\label{fig:nakamura}
\end{figure}
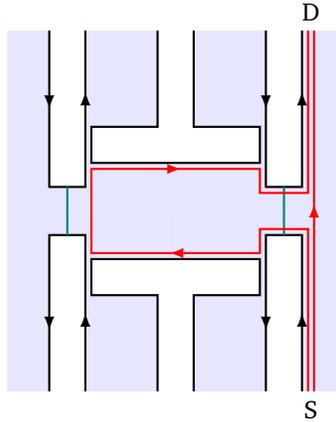

The experiment of Nakamura \etal\cite{nakamura-20np931,nakamura-22nc344} is a kind of Fabry--Perot interferometer.   It central idea is depicted schematically in Figure~\ref{fig:nakamura}.  As in Figure \ref{fig:feve}, the blue area depicts a $\nu = 1/3$ quantum Hall liquid, the white areas define contacts, S and D are anyon source and detector, respectively, the short green lines show weak links, and the red curves show paths from D to S that pass through one weak link.  The contacts circumscribe an island of quantum Hall liquid.  One of the paths encircles the island, while the other does not.  The relative phase between contributions from those paths depends continuously on the magnetic flux enclosed, and discretely on the number of bulk anyons in the island.  As one varies the applied magnetic field, discrete phase slips of appropriate magnitude (i.e., $2\pi/3$) are observed in the current interference pattern, corresponding to the creation of an additional anyon in the island,   This is quite direct evidence for the existence of anyons and their characteristic braiding behavior.  

A key innovation in this experiment, which allowed it to succeed where many earlier attempts had failed, was the addition of parallel electron gases that helped to screen long-range Coulomb forces within the active experimental layer.  This allowed the effect of the quantum statistical interaction, which is not screened, to emerge more clearly.   This technique will certainly empower similar measurements at other fractions, where more complex edge structures (featuring a spectrum of anyons)  and non-Abelian statistics might come into play.   


The model realization of unconventional quantum statistics in engineered systems has been pioneered by Chinese scientists.   The leading idea in this work is to capture the geometry of the mapping from positions of configurations of anyons in real space to states in Hilbert space that one has in an anyon model within truncated versions of both those spaces that can be realized using modern quantum technology.  This strategy has been implemented successfully on several platforms: quantum optics\cite{lu-09prl030502}, cold atoms\cite{dai-17np1195}, and superconducting circuits\cite{song-18prl030502}.   The details are ingenious, and this subject is clearly open-ended, as we will discuss further below.


Finally let us mention that Microsoft has put forward an ambitious roadmap for scalable topological computing\cite{karzig-17prb235305}, based on ising anyons (Majorana modes) realized at the ends of superconducting wires\cite{kitaev01pusp131}.  Recently they have reported evidence for the existence of the modes, but not yet for their predicted braiding properties.

\section{Summary and future directions }

In the bulk of this review we have sought to provide a self-contained introduction to anyon physics, emphasizing basic theoretical concepts and mature experimental examples.  

Before closing, let us briefly indicate a few directions in which future research is likely to be fruitful:

\begin{itemize}

\item {\it States of matter with anyonic quasiparticles: exotic superconductors, fractional Chern insulators, spin liquids---and more?}

The fractional quantum Hall effect is the most well-understood and well-established arena for anyon physics, but there are several other conjectured states of matter in 2+1 dimensions that, if they do exist, are predicted to host quasiparticles with unconventional quantum statistics.  These include, in addition to the p-wave superconductors mentioned earlier, fractional Chern insulators\cite{regnault-11prx021014} and several varieties of spin liquids\cite{savary-16rpp016502}, notably including some tractable spin models proposed by Kitaev\cite{kitaev03ap2}.  
More broadly: Within the framework of Landau-Ginzburg theory, it is natural to consider models that couple conserved currents (including topological currents) to gauge fields governed by Chern-Simons terms.  This construction will implement statistical transmutation for the particles that carry the associated charges.  The resulting models embody the general principles of quantum theory and locality, and can readily incorporate a wide variety of appropriate symmetries such as Galilean invariance, rotation symmetry, spin rotation symmetry, electron or other number conservation laws, and even $P$ or $T$ (see below), while postulating only a small number of emergent degrees of freedom.   Thus, they define plausible universality classes for the description of real materials whose excitations obey unconventional quantum statistics.

The question arises, of course, whether a candidate real material actually embodies such a universality class, and how to show that experimentally.  Here Wigner's insight \cite{wigner48pr1002} that the energy dependence of production cross-sections near their thresholds is strongly affected  on relative orbital angular momentum in the final state.  Indeed, the centrifugal barriers strongly influence the behavior of wave functions at small wave-vectors, and thus access to the (short-range) interaction region.  This effect is directly reflected in calculable power-law behavior.  Since fractional statistics is associated to fractional angular momentum, measurements using this effect can provide precise measures of quantum statistics, including mutual statistics\cite{morampudi-17prl227201}.  

Another characteristic feature of models of this kind is the occurrence, in bounded samples, of edge modes.  Theoretically, these arise because the gauge invariance of the Chern-Simons term depends on an integration by parts, and threatens to be spoiled by a surface term.  The surface term can be cancelled, however, by the contribution of an anomalous theory localized on the edge.  Thus, an  anomalous edge theory should be considered as part of the Chern-Simons construction.  This edge theory is necessarily chiral, and contains massless modes. 

Worthy of explicit mention is the fact that it is possible to have anyon quasiparticles even if the discrete symmetries P and T are unbroken.  Thus, for example, if our theory contains two different species of (anyonic) quasiparticles with equal and opposite values of $\theta$ but otherwise identical properties, then the modified \mbox{P} or T operations that combine the naive space or time reflections with interchange of those two species will be valid symmetries of the theory.

\item {\it States of anyon matter}

The states of matter that support anyon quasiparticles can in principle support a finite density of them.  The question thereby arises, how to describe the behavior of ensembles of anyons.  

Anyons in fractional quantum Hall states have special features that drastically affect their multi-particle dynamics.  Notably, since they are fractionally charged particles subject to a strong magnetic field, their kinetic energy is quenched and they can exert Coulomb interactions upon one another.  (In principle screening can ameliorate these effects, but in practice they are usually quite important.  Another complication is that charged anyons can be pinned by impurities.)    Still, fractional statistics plays a key role in the hierarchical construction of fractional quantum Hall states through iterated condensation of anyon ensembles.  

Statistical transmutation and the powerful 
adiabatic or heuristic principle
\cite{greiter-90mplb1063,hansson-21proc,greiter-21prbL121111} provide an alternative conceptual framework in which ensembles of anyons, generally subject to a uniform magnetic field, systematically interpolate between ensembles of fermions (that can be identified with electrons) or bosons.   Here the leading idea is that Chern-Simons flux, localized on particles, can be approximated by an equal amount of uniformly distributed flux.  This idea also unifies the description of discretely different purely phases of fermionic or bosonic matter (\eg fractional quantum Hall states with different fractions).  They come to be seem as continuously related conceptually---and perhaps physically---through intermediate states of anyon matter.

Unfortunately, there seems to be no usable analogue, for anyons, of the independent particle approximation that proves so useful for fermions and bosons.  That is, one cannot construct candidate wave-functions for multi-anyon systems by taking appropriately symmetrized products of single-particle wave functions.   Altogether, there seems to be ample room for creativity and experimentation in the future exploration of possible states of anyon matter. 

\item {\it Encoding and processing quantum information} 


The statistical interaction among anyons is governed by the topology of their world lines.  This implies that this interaction, and its action on many-anyon wave functions, is unaffected by small perturbations of the particle paths, or---to put it in more suggestive terms, that the quantum information encoded and sensed by anyons is distributed globally, rather than locally.   This feature renders that information insensitive to small local perturbations, and thus to many forms of noise.  Since the fragility of quantum information is a major challenge inhibiting its practical use, there is great interest in using anyons, or the ideas around them, for quantum information engineering.  

To be more concrete, consider for example anyons on a torus.  A flux line of the statistical gauge field that wraps around a cycle of the torus is hard to destroy, but it can be ``read out''---that is, detected---by anyons that move through the other cycle.  This set-up inspired the toric code construction of robust qubits\cite{kitaev06ap2}, elaborations of which feature in currently cutting-edge quantum computing hardware\cite{nigg-14s302}. 

The practical implementation of this line of thought in physical circuitry brings in many new questions and opportunities, since the most natural description of the circuity will usually start off looking quite different from particles and flux tubes, not to mention Chern-Simons field theories.  Here too, there is ample scope for further creativity.  

Let us indicate a framework that retains the conceptual essence of anyon physics while potentially allowing much more flexible embodiment.  In anyon physics, as described above, we have some number $N$ of particles with distinct positions $x^j$ on a two-dimensional surface.  (For concreteness and simplicity of notation we 
assume one species of anyon, but generalization to many species is straightforward.)  Associated with these configurations is a family of Hilbert spaces ${\cal H} (x^j)$ depending continuously on the $x^j$, all of the same dimension---\viz the space of degenerate ground states in the presence of the anyons---generally realized as subspaces of a master Hilbert space ${\cal H}$.  And we have Hermitian evolution operators $A_k (x^j)$ whose path-ordered exponentiation leads to unitary transformations
\begin{equation}\label{eq:master_connection}
U(\Gamma; x_F, x_I) \equiv \text{P} \exp \, \biggl(\, i \int \, d\lambda \ A_k (x(\lambda)) \, \frac{dx^k}{d\lambda} \, \biggr)
\end{equation}
which map ${\cal H} (x_I)$ to ${\cal H}(x_F)$ and are functions of parameterized trajectories $\Gamma$ connecting an initial configuration $x_I$ to a final configuration $x_F$.

Note that nothing in this set-up requires the $x^k$ to be positions of particles or that the master Hilbert space should be the Hilbert space of a continuum quantum field theory.  For example, the $x^j$ might be parameters characterizing adjustments to a quantum circuit and the $\cal H$ the state space of a finite number of qubits.  Indeed, since \eqref{eq:master_connection} governs the adiabatic evolution of quantum systems in response to variation of parameters, with $A_k$ the (generally non-Abelian) geometric or Berry phase, one can imagine a wide variety of realizations.  From this point of view, however, the topological character of the result---that is, the invariance $U(\Gamma; x_F, x_I)$  to small perturbations of $\Gamma$ that leave the end-points fixed---is not generic.   It requires that $A$, regarded as a gauge field on the parameter space, has zero curvature.   Thus, in order to preserve the advantages of anyons for noise immunity within this more general construction one should avoid visiting regions of large curvature.


\item {\it Topological interactions of extended objects}

Finally, let us briefly mention an extension of the circle of ideas around fractional statistics that take us beyond the realm of quantum-mechanical particles.  

As mentioned just above, the Lagrangian Chern-Simons construction can be carried out with the gauge field coupling to currents associated to field topology, \eg vorticity or Skyrmion number.  This possibility exists even in the realm of classical field theory\cite{radu-08prep101}.  The dynamics of winding far-separated concentrations of the topological charge around one another will reflect the ``statistical phase'' as a real-valued contribution to the action. Such interactions affect the equations of motion, and can alter the dynamics in interesting ways\cite{garaud-22jhep1029}, for example by lowering the energy for one direction of orbital motion relative to the other.   In principle (and in practice, see reference \cite{HasegawaKodama95}) arrays of classical solitons can be used to encode and transmit information, and the topological character of these interactions might be helpful in ameliorating the effects of noise in this context too.

\end{itemize}



\subsection*{Acknowledgements}

We wish to thank Tobias Helbig and Tapash Chakraborty for helpful suggestions on the manuscript.  MG is supported by the Deutsche Forschungsgemeinschaft (DFG, German Research Foundation)---Project-ID 258499086---SFB 1170, through the Würzburg-Dresden Cluster of Excellence on Complexity and Topology in Quantum Matter---\textit{ct.qmat} Project-ID 390858490---EXC 2147, and through DFG grant GR-1715/3-1.  FW is supported by the U.S. Department of Energy under grant Contract Number DE-SC0012567, by the European Research Council under grant 742104, and by the Swedish Research Council under Contract No. 335-2014-7424.

\medskip

\renewcommand{\refname}{\normalsize Further Reading}

%
%
%

\end{document}